\documentclass{article}

\usepackage{latexsym}
\usepackage{amsmath}
\usepackage{amssymb}

\usepackage{graphicx}
\usepackage{subfigure}

\newcommand{\pz}{\sqrt{e^{2\phi}+|p|^2}}
\newcommand{\pzn}{\sqrt{e^{2\phi_n}+|p|^2}}
\newcommand{\pzo}{\sqrt{e^{2\phi_0}+|p|^2}}
\newcommand{\R}{\mathbb{R}}
\def\prfe{\hspace*{\fill} $\Box$

\smallskip \noindent}

\newtheorem{theorem}{Theorem}
\newtheorem{proposition}{Proposition}
\newtheorem{lemma}{Lemma}
\newtheorem{remark}{Remark}
\newtheorem{corollary}{Corollary}

\begin{document}

\title{Asymptotic behavior  \\ and  orbital stability of galactic dynamics  \\ in relativistic 
scalar gravity}
\author{Simone Calogero$^{\: a}$\footnote{calogero@mct.uminho.pt}\,\,,\ 
\'Oscar S\'anchez$^{\: b}$\footnote{ossanche@ugr.es}\, and Juan 
Soler$^{\: b}$\footnote{jsoler@ugr.es}\\[0.5cm]
$^a$Departamento de Matem\'atica para a Ci\^encia e Tecnologia\\
Campus de Azur\'em da Universidade do Minho\\
4800-058 Guimar\~aes, Portugal\\[0.3cm]
$^b$Departamento de Matem\'atica Aplicada\\
Facultad de Ciencias, Universidad de Granada\\
18071 Granada, Spain}\date { }

\maketitle

\begin{center}
{\it 
Dedicated to Juan Luis V\'azquez on his 60th birthday}
\end{center}
\begin{abstract}
The Nordstr\"om-Vlasov system is a relativistic Lorentz invariant generalization of the Vlasov-Poisson system in the gravitational case. The asymptotic behavior of solutions and the non-linear stability of steady states are investigated. It is shown that solutions of the Nordstr\"om-Vlasov system with energy grater or equal to the mass satisfy a dispersion estimate in terms of the conformal energy. When the energy is smaller than the mass, we prove existence and non-linear (orbital) stability of a class of static solutions (isotropic polytropes) against general perturbations. The proof of orbital stability is based on a variational problem associated to the minimization of the energy functional under suitable constraints.
\end{abstract}

\section{Introduction}\label{intro}
A classical problem in theoretical astrophysics is to establish the 
non-linear stability of galaxies in equilibrium.  Neglecting 
relativistic effects and collisions among the stars of the galaxy, 
these equilbrium states can be described as stationary solutions of 
the Vlasov-Poisson system:
\begin{eqnarray*}
&&\partial_t f+p\cdot\nabla_x f-\nabla_x \phi\cdot\nabla_p f=0,\\
&&\bigtriangleup_x \phi=4\pi\rho,\ \lim_{|x|\to\infty} \phi(t,x)=0,\\
&&\rho(t,x)=\int_{\R^3} f(t,x,p)\,dp.
\end{eqnarray*}
Here, $f(t,x,p)\geq0$ denotes the distribution function in 
phase space of the stars, which are assumed to have all the same mass, with $t\in\R$, $x\in\R^3$, $p\in\R^3$ denoting time, position and momentum, respectively; 
$\phi(t,x)$ stands for the mean gravitational 
field generated by the stars altogether. In our units, the 
gravitational constant and the mass of each star equal one.

A general method to approach the stability problem for an infinite 
dimensional dynamical system is to construct stationary solutions  as 
minimizers of a suitable functional which is preserved by the 
evolution. The specific choice of the functional to minimize and of 
the constraints in the variational problem selects the type of steady 
states to be constructed, as well as the notion of distance which 
appears in the non-linear stability theorem. This approach was 
successfully applied to establish non-linear stability for a large 
class of steady states to the Vlasov-Poisson system, see \cite{DSO,G1,G2,GR1,GR2,R1,R2,R3,SS,W}. An important pre-requisite for 
this method to yield a rigorous stability theorem is that 
sufficiently regular solutions of the dynamical system should exist 
for general initial data of the Cauchy problem (or at least for data 
close to the steady state). In the case of Vlasov-Poisson, the 
existence of global classical solutions for general initial data has 
been known for some time, see \cite{LP,Pf,R3,Sch}.

In this paper we study  the non-linear stability for a class of stationary solutions as well as the asymptotic behaviour of general solutions to the Nordstr\"om-Vlasov system  
\cite{C,CL}. The latter provides a genuine relativistic generalization of the Vlasov-Poisson system in the following sense: It is invariant under Lorentz transformations and its solutions converge to solutions of Vlasov-Poisson as the speed of light tends to infinity. In units such that the speed of light equals one, the Nordstr\"om-Vlasov system is given by
\begin{equation}\label{nordstrom}
\partial_t^2\phi-\bigtriangleup_x\phi=-\int_{\R^3} 
f(t,x,p)\,\frac{dp}{\sqrt{1+|p|^2}},
\end{equation}
\begin{equation}\label{vlasov}
Sf
- \left[(S\phi)\,p + (1+|p|^2)^{-1/2} \nabla_x\phi 
\right]\cdot\nabla_p f= 4 f\, S\phi,
\end{equation}
where
\[
\widehat{p}=\frac{p}{\sqrt{1+ |p|^2}},\quad 
S=\partial_t+\widehat{p}\cdot\nabla_x
\]
are the relativistic velocity and the relativistic free transport operator.
As compared to the Vlasov-Poisson case, the stability analysis for the Nordstr\"om-Vlasov system presents new difficulties related to the strongly nonlinear 
and hyperbolic character of the field equations. These new features make the problem under study relevant from a mathematical point of view, besides its original astrophysical motivation.

A fundamental role in our analysis is played by the conserved quantities for the system (\ref{nordstrom})-(\ref{vlasov}).
Let us mention here the energy 
functional, or Hamiltonian,
\[
H(f,\phi,\partial_t\phi)=\int_{\R^3}\int_{\R^3}\sqrt{1+|p|^2}\,f\,dp\,dx+\frac{1}{2}\int_{\R^3}|\nabla_x\phi|^2dx+\frac{1}{2}\int_{\R^3}(\partial_t\phi)^2dx
\]
and the Casimir functional
\[
\mathcal{C}_Q(f,\phi)=\int_{\R^3}\int_{\R^3} e^{3\phi}Q(fe^{-4\phi})\,dp\,dx,
\]
where $Q:\R\to\R$ is any sufficiently regular function. In 
particular, for $Q(z)=z^q$, $q\geq1$, we infer that
$\|e^{(3/q-4)\phi} f\|_{L^q}$ is constant,
the case $q=1$ being the conservation law of mass. For the purpose of 
the present investigation, it is convenient to introduce the new 
dynamical variable
\[
\widetilde{f}(t,x,p)=e^{-4\phi}f(t,x,e^{-\phi}p),
\]
in terms of which the Nordstr\"om-Vlasov system takes the form
\begin{equation} \label{vlasovt}
\partial_t \widetilde{f} +\frac{p}{\pz}\cdot\nabla_x\widetilde{f}- 
\nabla_x\left(\pz\right)\cdot\nabla_p \widetilde{f} = 0,
\end{equation}
\begin{equation}\label{wavet}
\partial_t^2\phi-\bigtriangleup_x\phi = - e^{2\phi}\int_{\R^3} 
\widetilde{f}(t,x,p)\,\frac{dp}{\pz}.
\end{equation}
The energy functional becomes
\begin{eqnarray}
H(\widetilde{f},\phi,\partial_t\phi)=E_{\rm 
kin}(\widetilde{f},\phi)+\frac{1}{2}\int_{\R^3}|\nabla_x\phi|^2\,dx+\frac{1}{2}\int_{\R^3}\left(\partial_t\phi\right)^2\,dx,
\label{energy}
\end{eqnarray}
where
\[
E_{\rm kin}(\widetilde{f},\phi)=\int_{\R^3}\int_{\R^3}\pz\,\widetilde{f}\,dp\,dx,
\]
while the Casimir functional reads
\[
\mathcal{C}_Q(\widetilde{f}\,)=\int_{\R^3}\int_{\R^3} Q(\widetilde{f}\,)\,dp\,dx.
\]
In particular $\|\widetilde{f}(t)\|_{L^q}$ is constant, for all $q\in 
[1,\infty]$. The main reason to adopt this new formulation of the 
Nordstr\"om-Vlasov system is that the field $\phi$ does not appear explicitly in 
the definition of the Casimir functional. Let us also notice that the 
particles distribution $\widetilde{f}$ and the gravitational 
potential $\phi$ in the energy functional of the Nordstr\"om-Vlasov 
system are independent variables, whereas for Vlasov-Poisson they are 
related by the Poisson formula $\phi=-\rho*\frac{1}{|x|}$. In the following we 
shall denote $\widetilde{f}$ simply by $f$ to lighten the notation.

One of the objectives of this paper is to underline the differences
between the Vlasov-Poisson system and the Nordstr\"om-Vlasov system 
related to the relativistic character of the latter model. We will then focus
our attention on the stability analysis to a simple class of steady 
states, namely the isotropic polytropes, which in the case of the 
Nordstr\"om-Vlasov system are defined as
\begin{equation}\label{polytropes}
f_0(x,p)=\left(\frac{E_0-E}{c}\right)_+^k,\quad E=\pzo.
\end{equation}
Here $k>-1$, $c>0$ and $E_0>0$ are constants, $E$ is the particles energy, 
$(\cdot)_+$ denotes the positive part. The existence of a maximum $E_0$ for the particles
energy is necessary for the steady state to have 
finite energy \cite{C}.
Moreover, $\phi_0=\phi_0(x)$ is the gravitational potential induced 
by the distribution $f_0$ and is a solution of the non-linear Poisson 
equation
\begin{equation}\label{Poisson}
\bigtriangleup\phi_0=e^{2\phi_0}\int_{\R^3}\frac{f_0}{\pzo}\,dp.
\end{equation}

Our proof of stability for solutions of the form 
(\ref{polytropes})-(\ref{Poisson}) is grounded on a variational 
argument similar to the one introduced in \cite{SS} to study the orbital stability of 
polytropic spheres for the Vlasov-Poisson system and which consists 
in minimizing the energy functional subject to the constraints of 
given mass and $L^q$ norm of $f$, for some $q>1$.

This paper is organized as follows. We begin in Section \ref{dispersion} by proving that solutions of the Nordstr\"om-Vlasov system with energy greater or equal to the mass satisfy a dispersion estimate in terms of the conformal energy. This estimate does not imply that steady states solutions must have energy smaller than the mass, because for these solutions the conformal energy is unbounded. In Section \ref{results} we state our main results on the stability of isotropic polytropes for the Nordstr\"om-Vlasov system and reduce the problem to that of minimizing the energy functional under suitable constraints. Some preliminary estimates necessary to solve the latter problem are given in Section \ref{bounds}. In particular it is shown that the energy of minimizers is bounded above by their mass, thus a connection with the result of Section \ref{dispersion} is established. In Section \ref{existencemin} we prove the existence of minimizers to the energy functional and show that they arise as the strong limit of suitable minimizing sequences. Our proof requires the minimizers to have energy {\it strictly} less than the mass, a property which is shown to be verified if the mass is sufficiently large. Finally, in Section \ref{properties} we establish uniqueness and spherical symmetry of the minimizer (up to a translation in space) and show that it is an isotropic polytrope solution  of the Nordstr\"om-Vlasov system with finite radius.

To conclude this introduction we remark that relativistic theories of 
gravity, although not physically correct, are often used as simplified models for General 
Relativity \cite{ST}. Moreover, scalar fields play a central role in 
modern
theories of classical and quantum gravity \cite{DEF}. The
physically correct relativistic model for self-gravitating
collisionless matter is the Einstein-Vlasov system, which is
discussed for instance in \cite{And}. Existence and finite radius 
property of steady states to the Einstein-Vlasov system have been 
studied in \cite{RR}, see also \cite{FHU}, but the question of their stability is 
currently open (see however \cite{W2}). We hope that the present work 
on the stability of steady states to the Nordstr\"om-Vlasov system 
may contribute to a better understanding of this important problem.

\section{A dispersion estimate}\label{dispersion}
The aim of this section is to prove a dispersion estimate for solutions of the 
Nordstr\"om-Vlasov system. { Although this estimate will not be used in the following sections, it reveals an interesting link with the assumptions in our stability theorem, see Section \ref{results}.}  
Define the local energy of a solution $(f,\phi)$ of (\ref{vlasovt})-(\ref{wavet}) as
\[
e(t,x)=\int_{\R^3} 
\sqrt{e^{2\phi}+|p|^2}\,f\,dp+\frac{1}{2}(\partial_t\phi)^2+\frac{1}{2}(\nabla_x\phi)^2,
\]
the local momentum
\[
q_i(t,x)=\int_{\R^3} p_i\,f\,dp-\partial_t\phi\partial_i\phi,
\]
and the local stress tensor
\[
\tau_{ij}=\int_{\R^3} \frac{p_i\,p_j}{\sqrt{e^{2\phi}+|p|^2}}\,f\,dp+\partial_i\phi\,\partial_j\phi+\frac{1}{2}\delta_{ij}\left[(\partial_t\phi)^2-(\nabla_x\phi)^2\right],
\]
where $\partial_i=\partial_{x_i}$. These quantities are
related by the conservation laws
\begin{equation}\label{localcons}
\partial_te+\nabla_x\cdot q=0,\quad\partial_t q_i+\partial_j\tau_{ij}=0,
\end{equation}
the sum over repeated indexes being understood.
Upon integration, the previous identities lead to the conservation of 
the total energy and of the total momentum:
\[
H(t)=\int_{\R^3} e\,dx=constant,\quad Q(t)=\int_{\R^3} q\,dx=constant.
\]
Moreover, solutions of (\ref{vlasovt})-(\ref{wavet}) satisfy the 
conservation of the total rest mass:
\[
M=\int_{\R^3} f\,dp\,dx=constant,
\]
which is obtained by integrating the local rest mass conservation law
\[
\partial_t\rho+\nabla_x\cdot j=0,\quad \rho=\int_{\R^3} f\,dp,\quad 
j=e^{\phi}\int_{\R^3} \frac{p}{\sqrt{e^{2\phi}+|p|^2}}\,f\,dp.
\]
We define the conformal energy as
\[
\mathcal{E}_C(t)=\int_{\R^3} |x|^2e(t,x)\,dx.
\]
The Nordstr\"om-Vlasov system is supplied with a set of initial data 
$(f_0,\phi_0,\phi_1)$, where
\[
f_0(x,p)=f(0,x,p),\quad \phi_0(x)=\phi(0,x),\quad\phi_1(x)=\partial_t\phi(0,x).
\]
In \cite{C2} it is proved that $f_0\in C^1_c,\,\phi_0\in C^3_b\cap 
H^1,\,\phi_1\in C^2_b\cap L^2$ launch a unique global classical 
solution of (\ref{vlasovt})-(\ref{wavet}). This solution preserves 
energy and mass and has finite conformal energy for all times (provided it is bounded at time zero, {\it e.g}, by giving compactly supported initial data for the field).
We prove the following

\begin{theorem}\label{maintheo}
There exists a constant $t_0>0$, depending only on bounds on the 
initial data, such that the following holds:
If the initial data satisfy $H>M$ then
\[
\mathcal{E}_C(t)\geq(H-M)\, t^2,
\]
for all $t>t_0$; if $H=M$, then
\[
\mathcal{E}_C(t)\geq2Q_0\, t,
\]
again for $t>t_0$, provided $Q_0>0$, where
\[
Q_0=\int_{\R^3} (x\cdot q(0,x)-\phi_0\phi_1)\,dx.
\]
\end{theorem}
\noindent\textit{Proof:}  By the first of (\ref{localcons}) we have
\begin{equation}\label{firstderE}
\frac{d}{dt}\mathcal{E}_C=2\int_{\R^3} x\cdot q\,dx,
\end{equation}
whence, using the second equation in (\ref{localcons}),
\begin{equation}\label{secondderE}
\frac{d^2}{dt^2}\mathcal{E}_C=2\int_{\R^3}{\rm Tr}(\tau_{ij})\,dx.
\end{equation}
Here ${\rm Tr}(\tau_{ij})$ denotes the trace of the tensor 
$\tau_{ij}$ which is given by
\[
{\rm Tr}(\tau_{ij})=\int_{\R^3} \frac{|p|^2}{\sqrt{e^{2\phi}+|p|^2}}\,f\,dp-\frac{1}{2}(\nabla_x\phi)^2+\frac{3}{2}(\partial_t\phi)^2.
\]
It follows that one can rewrite (\ref{secondderE}) as
\begin{equation}\label{secondderE2}
\frac{d^2}{dt^2}\mathcal{E}_C=2H+2\mathcal{Q}(\partial_t\phi,\nabla_x\phi)-2\int_{\R^3} \mu(t,x)\,dx,
\end{equation}
where $\mu$ is minus the right hand side of (\ref{wavet}) and $\mathcal{Q}$ is the quadratic operator
\[
\mathcal{Q}(\partial_t\phi,\nabla_x\phi)=\int_{\R^3}\left((\partial_t\phi)^2-(\nabla_x\phi)^2\right)\,dx.
\]
By means of the identity 
$(\partial_t\phi)^2=\partial_t(\phi\partial_t\phi)-\phi\partial_t^2\phi$ 
and using (\ref{wavet}) we have
\begin{equation}\label{identity}
\int_0^t\mathcal{Q}(\partial_t\phi,\nabla_x\phi)\,ds=\int_0^t\int_{\R^3}\mu\,\phi\,\,dx\,ds+\frac{1}{2}\partial_t\int_{\R^3}\phi^2\,dx-\int_{\R^3}\phi_0\phi_1\,dx.
\end{equation}
From (\ref{firstderE}), (\ref{secondderE2}) and (\ref{identity}) we obtain
\begin{eqnarray}
\mathcal{E}_C(t)&=&\mathcal{E}_C(0)-\int_{\R^3}\phi_0^2\,dx+\int_{\R^3}\phi^2\,dx+2Q_0t+Ht^2\nonumber\\
&&+2\int_0^t\int_0^s\int_{\R^3} \mu\,(\phi-1)\,dx\,d\tau\,ds\label{identity2}.
\end{eqnarray}
Using the simple lower bound 
$\xi-1\geq-e^{-\xi}$, which holds for all $\xi\in\R$, the last term in (\ref{identity2}) is bounded from below by 
\[
-2\int_0^t\int_0^s\int_{\R^3} 
\int_{\R^3} \frac{f\,e^{\phi}}{\sqrt{e^{2\phi}+|p|^2}}\,dp\,dx\,d\tau\,ds\geq-Mt^2.
\]
Substituting into (\ref{identity2}) we finally obtain
\[
\mathcal{E}_C(t)\geq\mathcal{E}_C(0)-\|\phi_0\|_{L^2}^2+2Q_0t+(H-M)t^2,
\]
which yields the claim.\prfe

\begin{remark}
\textnormal{The estimate of Theorem \ref{maintheo} does not apply to steady states solutions. To see this, consider $f_0$ with compact support and $\phi_0$ a solution of (\ref{Poisson}). The fastest decay at infinity for $\nabla\phi_0$ is $O(|x|^{-2})$, which is too weak to bound the integral $\int_{\R^3} |x|^2|\nabla\phi_0|^2dx$ in the conformal energy. Nevertheless the bound $H<M$ will also appear as a crucial ingredient in our proof of stability for isotropic polytropes.}
\end{remark} 

\section{Orbital stability}\label{results}
In this section we state and comment our main results on the stability of isotropic polytropes of the Nordstr\"om-Vlasov system.
In order to be precise with the formulation of our problem, { we will start by introducing some notation.}
{ For $J$, $M$, $k$ positive real numbers, we denote by $\Gamma_{M,J}^k$ the space of functions} $f:\R^6\to\R$ given by 
\[
\Gamma_{M,J}^k = \{ f\in L^1\cap L^{1+1/k},\ f\geq0\ \textnormal{a.e.} ,\ \|f\|_{L^1}=M,\ \|f\|_{L^{1+1/k}}\leq J \}.
\]
Moreover we denote
\[
D^1(\R^3)=\{\phi\in L^1_{\rm loc}(\R^3):\,\nabla\phi\in 
L^2\textnormal{ and }\phi \textnormal{ vanishes at infinity}\},
\]
where the condition of $\phi$ vanishing at infinity means that the 
set $\{x\in\R^3:|\phi(x)|>a\}$ has finite (Lebesgue) measure, for all 
$a>0$.  Functions in the space $D^1(\R^3)$ satisfy the Sobolev inequality
\begin{equation}\label{sobolev}
\|\phi\|_{L^6}\leq 
\eta\|\nabla\phi\|_{L^2},\quad\eta=\frac{2}{\sqrt{3}}\pi^{-2/3},
\end{equation}
see \cite[Thm.~8.3]{LL}.
The space $D^1(\R^3)$ has been extensively used in fluid mechanics, see for example \cite{Li}. 

Our first result is the existence of a 
minimizer to the variational problem
\[
\inf\{H(f,\phi,\psi),\ f\in\Gamma_{M,J}^k,\ \phi\in D^1,\ 
\psi\in L^2,\ E_{\rm kin}(f,0)<\infty\},\quad k\in (0,2),
\]
provided the mass $M$ is sufficiently large (depending on $J$ and 
$k$), where $H$ is the energy functional defined in (\ref{energy}). Obviously, the above variational problem is equivalent to the 
following one:
\[
I_{M,J}^k=\inf\{\mathcal{E}(f,\phi),\ f\in\Gamma_{M,J}^k,\ \phi\in 
D^1,\ E_{\rm kin}(f,0)<\infty\},
\]
where
\begin{equation}\label{functional}
\mathcal{E}(f,\phi)=H(f,\phi,0)=E_{\rm 
kin}(f,\phi)+\frac{1}{2}\int_{\R^3}|\nabla\phi|^2dx.
\end{equation}
By abuse of terminology, we shall continue to refer to $\mathcal{E}$ 
as the energy functional.

\begin{theorem}\label{existence}
For all $0<k<2$ and $J>0$, there exists $M_0\in [0,\infty)$ such that 
$I_{M,J}^k<M$, for all $M>M_0$. Moreover, for all $M>M_0$ and
 any minimizing sequence 
$(f_n,\phi_n)\subset\Gamma_{M,J}^k\times D^1$ of the functional 
(\ref{functional}), there exist a subsequence, still denoted 
$(f_n,\phi_n)$, a sequence of spatial translations 
$T_nf_n(x,p)=f_n(x+y_n,p)$, $T_n\phi_n(x)=\phi_n(x+y_n)$, with 
$y_n\in\R^3$, and a minimizer $(f_0,\phi_0)\in\Gamma_{M,J}^k\times 
D^1$ such that
\[
\|T_nf_n-f_0\|_{L^1}\to 0,\ \|T_nf_n-f_0\|_{L^{1+1/k}}\to 0,\ \|\nabla 
T_n\phi_n-\nabla\phi_0\|_{L^2}\to 0,
\]
as $n\to\infty$.
Moreover, $(f_0,\phi_0)$ satisfies (\ref{Poisson}) in the sense of 
distributions.
\end{theorem}
Let us comment some aspects concerning the statements of Theorem \ref{existence}.
\begin{itemize}
\item[i)]  As in the case of Vlasov-Poisson, the use of translations in the 
space variable is necessary, otherwise starting from a minimizer 
$(f_0,\phi_0)$ and a sequence of shift vectors $y_n\in\R^3$ such that 
$\lim_{n\to\infty}|y_n|=\infty$, the sequence 
$(T_nf_0,T_n\phi_0)$---which is still minimizing---converges weakly 
to zero, which is not in $\Gamma_{M,J}^k\times D^1$.
\item[ii)]  With regard to the large mass condition, we notice first that if 
$M_0=0$, then our results apply to any $M>0$. If $M_0>0$, then $I_{M,J}^k=M$, for all $0<M<M_0$ and in this case it is possible to construct a minimizing sequence of 
the functional $\mathcal{E}$ with vanishing weak limit, see Remark 
\ref{largemasscondition} in Section \ref{bounds}. 
\item[iii)] { In view of Theorem \ref{maintheo} and the previous remark, it is reasonable to conjecture that the condition $I_{M,J}^k<M$ in Theorem \ref{existence} is optimal.} 
\end{itemize}

As an application of the Euler-Lagrange multipliers method we show 
that minimizers are isotropic polytropes solutions to the 
Nordstr\"om-Vlasov system.

\begin{theorem}\label{Lagrange}
Let $0<k<2$, $J>0$, $M>M_0$ and $(f_0,\phi_0)$ be a minimizer of 
(\ref{functional}) over the space $\Gamma_{M,J}^k\times D^1$. Then 
$f_0$  has the form (\ref{polytropes}), with 
\[
E_0 \in \left(\frac{k+4}{6} \frac{I_{M,J}^k}{M},\frac{I_{M,J}^k}{M}\right),\quad c>0 
\]
given by
\[
E_0=\frac{1}{M}\left(I_{M,J}^k-\frac{2-k}{6}\int_{\R^3}|\nabla\phi_0|^2dx\right),\quad
c=\frac{k+1}{2-k}\left(\frac{I_{M,J}^k-E_0M}{J^{1+1/k}}\right).
\]
Moreover different minimizers differ only by a spatial translation and for any representative in this class the following holds:  $\phi_0(x)\to 0$, as $|x|\to\infty$, $f_0(x,p)$ is compactly 
supported, $(f_0,\phi_0)$ is spherically symmetric with respect to 
some point in $\R^3$ and is a time-independent mild solution of 
the Nordstr\"om-Vlasov system.
\end{theorem}

Uniqueness of minimizers in Theorem \ref{Lagrange} (up to 
translations in space) is proved by showing that the Lagrange 
multiplier $E_0$ is the same for all minimizers. In the case of 
Vlasov-Poisson, this follows by the scaling properties of the 
Emden-Fowler equation, which is the ordinary differential equation (ODE) satisfied by the (spherically 
symmetric) gravitational potential of isotropic polytropes, see 
\cite{GR1}. Here, this argument does not apply due to the strongly nonlinear 
and nonlocal character of the ODE which is the equivalent
counterpart of the Emden-Fowler equation in the Nordstr\"om-Vlasov case.  In fact,
our strategy to prove uniqueness of minimizers for the 
Nordstr\"om-Vlasov system must combine analytical and numerical tools because
some integrals involved in the associated ODE cannot be explicitly calculated. Our numerical/analytical computations reveal a monotonicity property of the set of minimizers with respect to the mass, which yields uniqueness by the mass constraint.  

To conclude this section, we show how the previous results yield the orbital stability 
of the minimizer solution with respect to perturbed time-dependent solutions of the 
Nordstr\"om-Vlasov system. We adopt the expression {\it orbital stability} to invoke any
criterium for which the orbit, described by
 $\{(T_yf_0, T_y\phi_0)\,; y \in \R^3 \}$, of a
stationary  solution $(f_0, \phi_0)$ is the set of functions which remains close
to a perturbed Nordstr\"om-Vlasov  solution.
This concept has been widely used
in the literature, see \cite{SS} and the references therein.

\begin{theorem}\label{stability}
Let  $(f_0,\phi_0)$ be the minimizer associated to  $0<k<2$, $J>0$ and $M>M_0$.
For every $\varepsilon>0$, there exists 
$\delta=\delta(\varepsilon)$ such that, for all initial data $(f^{\rm 
in}, \phi_0^{\rm in},\phi_1^{\rm in})=(f,\phi,\partial_t\phi)_{|t=0}$ 
of the Nordstr\"om-Vlasov system in the class
\[
0\leq f^{\mathrm{in}}\in\Gamma_{M,J}^k\cap C^1_c,\ 
\phi_0^{\mathrm{in}}\in C^3\cap D^1,\ \phi_1^{\mathrm{in}}\in C^2\cap 
L^2
\]
and
\[
\left|H(f^{\mathrm{in}},\phi_0^{\rm 
in},\phi^{\rm in}_1)-H(f_0,\phi_0,0)\right|\leq\delta,
\]
the associated solution $(f,\phi)\in C^1\times C^2$ of 
(\ref{vlasovt})-(\ref{wavet}) satisfies, for all $t>0$,
\[
\inf_{y\in\R^3}\|f-T_yf_0\|_{L^1}+\inf_{y\in\R^3}\|f-T_yf_0\|_{L^{1+1/k}} \leq\varepsilon,
\]
\[
\inf_{y\in\R^3}\|\nabla\phi-T_y\nabla\phi_0\|_{L^2} +\|\partial_t\phi\|_{L^2}\leq\varepsilon.
\]
\end{theorem}
\noindent\textit{Proof: } If the thesis of Theorem \ref{stability} 
were false, there would exist $\varepsilon_0>0$, a sequence 
$(f_n^{\rm in},\phi_{0,n}^{\rm in}, \phi_{1,n}^{\rm in})$ and $t_n>0$ 
such that
\[
0\leq f_n^{\mathrm{in}}\in\Gamma_{M,J}^k\cap C^1_c,\ 
\phi_{0,n}^{\mathrm{in}}\in C^3\cap D^1,\ \phi_{1,n}^{\mathrm{in}}\in 
C^2\cap L^2,
\]
\[
\left|H(f_n^{\rm 
in},\phi_{0,n}^{\rm in}, \phi_{1,n}^{\rm 
in})-H(f_0,\phi_0,0)\right|<\frac{1}{n}
\]
and each of $\|f_n(t_n)-T_yf_0\|_{L^1},\ \|f_n(t_n)-T_yf_0\|_{L^{1+1/k}},\ 
\|\nabla\phi_n(t_n)-T_y\nabla\phi_0\|_{L^2}$ and 
$\|\partial_t\phi_n(t_n)\|_{L^2}$
is greater than $\varepsilon_0$
for all $y\in\R^3$ and $n\in\mathbb{N}$.  
Here 
$(f_n,\phi_n)$ is the solution of (\ref{vlasovt})-(\ref{wavet})
associated to the initial data set $(f_n^{\rm in},\phi_{0,n}^{\rm 
in}, \phi_{1,n}^{\rm in})$. On the other hand, by conservation of 
energy and $L^q$ norm, $(f_n(t_n),\phi_n(t_n))$ is a minimizing 
sequence. Hence the above conclusion contradicts the thesis of 
Theorem \ref{existence}.
\prfe

\section{Bounds on the energy infimum}\label{bounds}
Throughout the paper we denote by $C$ any positive constant that 
depends only on $M,J$ and $k$. Moreover, any subsequence of a 
minimizing sequence $(f_n,\phi_n)$ is still denoted $(f_n,\phi_n)$.
We start by collecting some important estimates on the functions
\[
\rho_{\! f}=\int_{\R^3} f\,dp,\quad \mu_{\! f}=\int_{\R^3} f\, \frac{dp}{|p|}
\]
induced by an element $f\in\Gamma_{M,J}^k$.
\begin{lemma}\label{estrho}
For $0<k<2$, let $\omega=3+k$, so that $1+1/\omega\in (6/5,4/3)$, and
\begin{equation}\label{j}
j=\frac{3}{2}+\frac{5-\omega}{2(\omega-1)}.
\end{equation}
Then for any $f\in\Gamma_{M,J}^k$, there exists a constant $C>0$ such that
\[
\int_{\R^3}\rho_{\! f}^{1+1/\omega}dx\leq C\left(\int_{\R^3} \int_{\R^3} 
f^{1+1/k}dp\,dx\right)^{k/\omega}\left(\int_{\R^3} \int_{\R^3}|p|\,f\,dp\,dx\right)^{3/\omega},
\]
\[
\int_{\R^3} \mu_{\! f}^{\ j}\,dx\leq C\left(\int_{\R^3} \int_{\R^3} 
f^{1+1/k}\right)^{\frac{2k}{\omega-1}}\left(\int_{\R^3} \int_{\R^3}|p|\,f\,dp\,dx\right)^{\frac{5-\omega}{\omega-1}},
\]
\[
\|\rho_{\! f}\|_{L^{6/5}}\leq C\left(\int_{\R^3} \int_{\R^3} 
f^{1+1/k}dp\,dx\right)^{k/6}\left(\int_{\R^3} \int_{\R^3}|p|\,f\,dp\,dx\right)^{1/2}.
\]
\end{lemma}
\noindent\textit{Proof: } For all $r\geq1$,  and $R>0$ we write
\begin{eqnarray*}
\int_{\R^3}\rho_{\! f}^{\ r}\,dx&=&\int_{\R^3}\left(\int_{|p|\leq 
R}f\,dp+\int_{|p|>R}f\,dp\right)^rdx\\
&\leq&C\int_{\R^3}\left[\left(\int_{\R^3} 
f^q\,dp\right)^{1/q}R^{3(q-1)/q}+\frac{1}{R}\int_{\R^3}|p|f\,dp\right]^rdx.
\end{eqnarray*}
Optimizing in $R$ and applying H\"older's inequality we obtain, for 
all $\alpha>1$,
\begin{eqnarray*}
\int_{\R^3}\rho_{\! f}^{\ r}dx&\leq& C\left[\int_{\R^3} \left(\int_{\R^3} 
f^q\,dp\right)^{\frac{r}{4q-3}}\left(\int_{\R^3}|p|f\,dp\right)^{\frac{3q-3}{4q-3}r}dx\right]\\
&\hspace{-2.4cm}\leq&\hspace{-1.2cm} C\left(\int_{\R^3} \left(\int_{\R^3} 
f^q\,dp\right)^{\frac{r\alpha}{4q-3}}\!\!dx\right)^{ \frac{1}{\alpha}} \left(\int_{\R^3} \left(\int_{\R^3}|p|f\,dp\right)^{\frac{3r\alpha(q-1)}{(4q-3)(\alpha-1)}}\!\!dx\right)^{ \frac{\alpha-1}{\alpha}}.
\end{eqnarray*}
Choosing $q=1+1/k$, $\alpha=\omega/k$ and $r=1+1/\omega$ yields the 
first estimate on $\rho_{\! f}$ and the one on $\mu_{\! f}$ is 
proved likewise. The bound on $\|\rho_{\! f}\|_{L^{6/5}}$ follows by 
interpolation with the finite mass constraint.
\prfe

We apply the preceeding estimates to show that the energy infimum is 
strictly positive.
\begin{lemma}\label{positive}
For all $0<k<2$, $J>0$ and $M>0$, $I_{M,J}^k>0$.
\end{lemma}
\noindent\textit{Proof: } Let $(f_n,\phi_n)\subset\Gamma_{M,J}^k\times 
D^1$ be a minimizing sequence. As $e^{\xi}\geq1+\xi$, for all 
$\xi\in\R$, we have the lower bound
\begin{eqnarray*}
E_{\mathrm{kin}}(f_n,\phi_n)&\geq& a\int_{\R^3} \int_{\R^3} 
e^{\phi_n}f_n\,dp\,dx+(1-a)\int_{\R^3} \int_{\R^3}|p|f_n\,dp\,dx\\
&\geq&aM+a\int_{\R^3}\phi_n\int_{\R^3} f_n\,dp\,dx+(1-a)\int_{\R^3} \int_{\R^3}|p|f_n\,dp\,dx,
\end{eqnarray*}
for all $a\in [0,1]$. Using H\"older's inequality, (\ref{sobolev}) 
and Lemma \ref{estrho} we find
\begin{eqnarray*}
E_{\mathrm{kin}}(f_n,\phi_n)&\geq& aM-a\|\phi_n\|_{L^6}\|\rho_{\!f_n}\|_{L^{6/5}}+C(1-a)\|\rho_{\! f_n}\|_{L^{6/5}}^2\\
&\geq&aM-a\eta\|\nabla\phi_n\|_{L^2}\|\rho_{\!f_n}\|_{L^{6/5}}+C(1-a)\|\rho_{\! f_n}\|_{L^{6/5}}^2,
\end{eqnarray*}
where the constant $C$ is independent from $n$. Letting 
$X_n=\|\nabla\phi_n\|_{L^2}$ and $Y_n=\|\rho_{\! f_n}\|_{L^{6/5}}$ we obtain
\begin{eqnarray*}
\mathcal{E}(f_n,\phi_n)&\geq& aM-a\eta 
X_nY_n+C(1-a)Y_n^2+\frac{1}{2}X_n^2\\
&\geq& 
aM+\frac{1}{2}(1-a\eta)X_n^2+\left[C(1-a)-\frac{1}{2}a\eta\right]Y_n^2.
\end{eqnarray*}
Choosing $a$ sufficiently small, precisely
\[
a\leq\min\left(\frac{1}{\eta},\frac{C}{C+\frac{1}{2}\eta}\right),
\]
we obtain $\mathcal{E}(f_n,\phi_n)\geq aM>0$. Note that $a<1$. 
Letting $n\to\infty$ concludes the proof.
\prfe

The proof of Theorem \ref{existence} requires precise bounds on 
$I_{M,J}^k$. We start with a simple scaling argument which relates 
the energy infimum for different masses.
\begin{lemma}\label{scaling}
For all $0<k<2$, $J>0$ and $0<M_1\leq M_2$,
\[
I_{M_2,J}^k\leq \frac{M_2}{M_1}I_{M_1,J}^k.
\]
\end{lemma}
\noindent\textit{Proof: } Let 
$(f_n,\phi_n)\subset\Gamma_{M_1,J}^k\times D^1$ be a minimizing 
sequence and define
\[
\widetilde{f}_n(x,p)=\alpha f_n(\beta 
x,p)\quad\widetilde{\phi}_n(x)=\phi_n(\beta x),
\]
where $\beta=M_1/M_2$ and $\alpha=\beta^2$. By direct computation,
\[
\|\widetilde{f}_n\|_{L^1}=M_2,\quad\|\widetilde{f}_n\|_{L^{1+1/k}}=\left(\frac{M_1}{M_2}\right)^{\frac{2-k}{1+k}}\|f_n\|_{L^{1+1/k}}\leq
J,
\]
so that 
$(\widetilde{f}_n,\widetilde{\phi}_n)\subset\Gamma_{M_2,J}^k\times 
D^1$. Moreover
\[
\mathcal{E}(\widetilde{f}_n,\widetilde{\phi}_n)=\frac{M_2}{M_1}\mathcal{E}(f_n,\phi_n),
\]
whence
\[
I_{M_2,J}^k\leq 
\lim_{n\to\infty}\,\mathcal{E}(\widetilde{f}_n,\widetilde{\phi}_n)=\frac{M_2}{M_1} 
I_{M_1,J}^k,
\]
which concludes the proof. 
\prfe

\begin{proposition}\label{boundinf}
The inequality $I_{M,J}^k\leq M$ holds, for all $0<k<2$, $J>0$ 
and $M>0$. Moreover, the strict inequality
\begin{equation}\label{largemass}
I_{M,J}^k<M
\end{equation}
holds if the mass $M$ is sufficiently large.
\end{proposition}
\noindent\textit{Proof: } Let $q=1+1/k>3/2$ and consider 
$f_\gamma\in\Gamma_{M,J}^k$ given by
\begin{equation}\label{vanishingsequence}
f_\gamma(x,p)=\left(\frac{J^q}{M}\right)^{\frac{1}{q-1}}\chi_{\{|x|\leq\beta\}}(x)\chi_{\{|p|\leq
\gamma\}}(p),
\end{equation}
where $\chi_A$ denotes the characteristic function of the set $A$ and
\[
\beta=\frac{1}{\gamma}\left[\left(\frac{M}{J}\right)^{\frac{q}{q-1}}\left(\frac{3}{4\pi}\right)^2\right]^{1/3}.
\]
Define $\phi_\alpha\in D^1$ as
\[
\phi_\alpha(x)=-\alpha\psi\left(\frac{x}{\beta}\right),
\]
where $\psi\in C_c^\infty(\R^3)$, $0\leq\psi\leq 1$, 
$\psi(y)\equiv 1$, for $|y|\leq 1$, $\psi(y)\equiv 0$, for 
$|y|\geq2$. We estimate the energy of $(f_\gamma,\phi_\alpha)$ 
as
\begin{eqnarray}
\mathcal{E}(f_\gamma,\phi_\alpha)&\leq&\int_{\R^3} \int_{\R^3} 
e^{\phi_\alpha}f_\gamma\,dp\,dx+\int_{\R^3} \int_{\R^3}|p|\,f_\gamma\,dp\,dx+\frac{1}{2}\int_{\R^3}|\nabla\phi_\alpha|^2dx\nonumber\\
&\leq&e^{-\alpha}M+\frac{3}{4}M\gamma+\alpha^2\beta 
K\label{temp1},
\end{eqnarray}
where, by (\ref{sobolev}),
\[
K=\int_{\R^3}|\nabla\psi|^2dx\geq\eta^{-2}\|\psi\|_{L^6}^2\geq
\left(\frac{3}{4}\right)^{2/3}\pi^{5/3}.
\]
In particular, for $\alpha=0$,
\[
I_{M,J}^k\leq\liminf_{\gamma\to 0}\,\mathcal{E}(f_\gamma,0)\leq M,
\]
which proves the first claim of the proposition. To prove the strict 
inequality (\ref{largemass}) for large masses, we optimize the 
estimate (\ref{temp1}) by choosing $\gamma=\gamma(\alpha)$ as
\[
\gamma=\sqrt{\frac{4K}{3M}}\left[\left(\frac{M}{J}\right)^{\frac{q}{q-1}}\left(\frac{3}{4\pi}\right)^2\right]^{1/6}\alpha.
\]
So doing we obtain the inequality
\begin{equation}\label{temp2}
\mathcal{E}(f_\alpha,\phi_\alpha)\leq M\left 
[e^{-\alpha}+A^{-1}\alpha\right],
\end{equation}
where $f_\alpha=f_{\gamma(\alpha)}$ and
\[
A=\frac{1}{\sqrt{K}}\left(\frac{4}{3}\right)^{5/6}\left(\frac{\pi}{8}\right)^{1/3}J^{\frac{q}{6q-6}}M^{\frac{2q-3}{6q-6}}.
\]
We choose $M$ so large so that $A^{-1}<1$. Precisely,
\[
M> 
\left[\left(\frac{3}{4}\right)^{5(q-1)}\left(\frac{8}{\pi}\right)^{2(q-1)}K^{3(q-1)}J^{-q}\right]^{1/(2q-3)}.
\]
Since $A^{-1}<1$, we can optimize (\ref{temp2}) by choosing 
$\alpha=\log A$. The inequality (\ref{temp2}) becomes
\[
\mathcal{E}(f_\alpha,\phi_\alpha)\leq A^{-1}\left(1+\log A\right)M<M,
\]
which concludes the proof of (\ref{largemass}). 
\prfe

\begin{remark}\label{largemasscondition}
\textnormal{The constant $M_0$ in the statement of Theorem \ref{existence} is defined as
\begin{equation}\label{infmass}
M_0=\inf\{M'>0:\,I_{M,J}^k<M, \textnormal{ for all } M> M'\}.
\end{equation}
By Proposition \ref{boundinf}, $M_0<\infty$. We were not able to show 
that $M_0=0$, i.e., that the strict inequality (\ref{largemass}) is 
always satisfied. If $M_0>0$, then it follows by Lemma \ref{scaling} 
that $I_{M,J}^k=M$, for all $0<M<M_0$. In this case, the sequence 
$(f_\gamma,0)$, with $f_\gamma$ given by (\ref{vanishingsequence}), 
is minimizing. Since this sequence converges weakly to zero, then 
Theorem \ref{existence} is false when $I_{M,J}^k=M$.}
\end{remark}

Further information on $I_{M,J}^k$ can be obtained by assuming that 
the infimum is achieved. For any function $g$, we denote by $g^*$ the 
non-decreasing symmetric rearrengement of $g$ with respect to the $x$ 
variable (see \cite{LL}).
\begin{lemma}\label{boundminimizer}
Let $(f_0,\phi_0)$ be a minimizer of the functional $\mathcal{E}$ 
over the space $\Gamma_{M,J}^k\times D^1$. Then
\begin{itemize}
\item[(i)]$\|f_0\|_{L^{1+1/k}}=J$ and $\phi_0\leq 0$ almost everywhere;
\item[(ii)] $(\phi_0, f_0)$ satisfy (\ref{Poisson}) in the sense of distributions;
\item[(iii)] The following identity is satisfied:
\[
\frac{1}{2}\int_{\R^3}|\nabla\phi_0|^2dx=\int_{\R^3} \int_{\R^3}\frac{|p|^2}{\pzo}f_0\,dp\,dx,
\]
which is the relativistic analogue of the Virial Theorem;
\item[(iv)] $(f_0^*,-\phi_0^*)$ is also a minimizer.
\end{itemize}
\end{lemma}
\noindent\textit{Proof: } Let $q=1+1/k>3/2$. Assume $\|f_0\|_{L^q}=K<J$ 
and let $\widetilde{K}\in(K,J)$. Define $\widetilde{f}(x,p)=\alpha^3 
f_0(\alpha x,p)$, $\widetilde{\phi}(x)=\phi_0(\alpha x)$, where 
$\alpha=(\widetilde{K}/K)^{q/(3q-3)}$.
In this way, $\|\widetilde{f}\|_{L^1}=M$, $\|\widetilde{f}\|_{L^q}=\widetilde{K}$ and
\begin{eqnarray*}
\mathcal{E}(\widetilde{f},\widetilde{\phi})=\int_{\R^3} \int_{\R^3}\pzo 
f_0\,dp\,dx  \hspace{2cm}\\+\frac{1}{2}\left(\frac{K}{\widetilde{K}}\right)^{\frac{q}{3q-3}}\int_{\R^3}|\nabla\phi_0|^2dx<\mathcal{E}(f_0,\phi_0),
\end{eqnarray*}
which is a contradiction to $(f_0,\phi_0)$ being a minimizer. If 
$\phi_0>0$ in a set of non-zero measure, then the pair 
$(f_0,-|\phi_0|)\in\Gamma_{M,J}^k\in D^1$ would have energy strictly 
less than $\mathcal{E}(f_0,\phi_0)$, which is again a contradiction. 
For the proof of (ii), let $\zeta\in C_c^\infty(\R^3)$. As 
$\phi_0+t\zeta\in D^1$, for all $t\in\R$, it has to be
\[
0=\left[\frac{d}{dt}\mathcal{E}(f_0,\phi_0+t\zeta)\right]_{t=0}=\int_{\R^6} \frac{e^{2\phi_0}}{\pzo}\,f_0\zeta\,d(p,x)+\int_{\R^3}\nabla\phi_0\cdot\nabla\zeta\,dx,
\]
which is the claim. Note that differentiation inside the integral in 
the kinetic energy is justified, since $\exp\phi_0$ is bounded (by 
one).
For the proof of the Virial Theorem, consider the uniparametric 
family of functions 
$(f_\alpha,\phi_\alpha)\subset\Gamma_{M,J}^k\times D^1$ given by
\[
f_\alpha(x,p)=f_0(\alpha x,\alpha^{-1}p),\quad \phi_\alpha(x)=\phi_0(\alpha x).
\]
Since $(f_0,\phi_0)$ is a minimizer, the equation
\[
\left[\frac{d}{d\alpha}\mathcal{E}\left(f_\alpha,\phi_\alpha\right)\right]_{\alpha=1}=0
\]
is to be satisfied, which is equivalent to identity claimed in the 
lemma. It remains to prove (iv). It follows by the general properties 
of symmetric rearrengements that 
$(f_0^*,-\phi^*_0)\in\Gamma_{M,J}^k\times D^1$. Moreover,
\begin{eqnarray}
\int_{\R^3}|\nabla\phi_0^*|^2dx\leq\int_{\R^3}|\nabla\phi_0|^2dx,
\label{41}
\end{eqnarray}
see \cite[Lemma 7.17]{LL}.
To reach our goal it is therefore enough to prove that
\begin{eqnarray}
\int_{\R^3} \int_{\R^3}\sqrt{e^{-2\phi^*_0}+|p|^2}f^*_0\,dp\,dx\leq 
\int_{\R^3} \int_{\R^3}\pzo f_0\,dp\,dx.
\label{42}
\end{eqnarray}
To this purpose we  use the layer cake representation of 
$\sqrt{e^{2x}+|p|^2}$. For $y>0$ we write
\[
\sqrt{e^{-2y}+|p|^2}=\int_y^\infty\frac{e^{-2s}}{\sqrt{e^{-2s}+|p|^2}}\,ds+|p|.
\]
Applying this to $y=-\phi_0$ we obtain
\begin{eqnarray*}
&& \hspace{-0.6cm}\int_{\R^3} \int_{\R^3}\pzo f_0\,dp\,dx 
\\&=&\int_{\R^3} \int_{\R^3}\left[\int_0^\infty\frac{e^{-2s}}{\sqrt{e^{-2s}+|p|^2}}\chi_{\{-\phi_0\leq
s\}}ds\right]\,f_0\,dp\,dx
+\int_{\R^3} \int_{\R^3}|p|\,f_0\,dp\,dx.
\end{eqnarray*}
Next we use that
\[
\int_{\R^6} \frac{e^{-2s}}{\sqrt{e^{-2s}+|p|^2}}\chi_{\{-\phi_0\leq 
s\}}f_0\,d(p,x)\geq
\int_{\R^6} \frac{e^{-2s}}{\sqrt{e^{-2s}+|p|^2}}\chi_{\{\phi^*_0\leq 
s\}}f^*_0\,d(p,x),
\]
see \cite{LL}, eq. (3), pag. 83, and then
\begin{eqnarray*}
\int_{\R^3} \int_{\R^3}\pzo 
f_0\,dp\,dx&\geq&\int_{\R^3} \int_{\R^3}\sqrt{e^{2\phi_0^*}+|p|^2}\,f_0^*\,dp\,dx\\
&\geq&\int_{\R^3} \int_{\R^3}\sqrt{e^{-2\phi_0^*}+|p|^2}\,f_0^*\,dp\,dx.
\end{eqnarray*}
This concludes the proof that $(f^*_0,-\phi^*_0)$ is a minimizer.
\prfe

\begin{remark}\textnormal{We shall prove in Section \ref{properties} 
that $(f_0,\phi_0)$ coincides with $(f_0^*,-\phi_0^*)$, up to a 
translation in space. In particular, minimizers are spherically 
symmetric with respect to some point in $\R^3$.}
\end{remark}


\section{Existence of a minimizer}\label{existencemin}
In this section we prove Theorem \ref{existence}. We split the proof 
in several lemmas. In view of Lemma \ref{boundminimizer} (i), it is 
necessary to show that the positive part of $\phi_n$ vanishes in the 
limit, which is done in the next lemma. Denote by $g^+=\max(g,0)$ and 
$g^-=\min(g,0)$, the positive and negative part of a real valued 
function $g$, respectively.
\begin{lemma}\label{phinegative}
Let $(f_n,\phi_n)$ be a minimizing sequence. Then $(f_n,\phi_n^-)$ is 
also a minimizing sequence and
\begin{equation}\label{samelimit1}
\|\nabla\phi_n-\nabla\phi^-_n\|_{L^2}\to 0,\quad n\to\infty.
\end{equation}
In particular, after possibly extracting a subsequence, $\phi^+_n\to 0$ a.e.
\end{lemma}
\noindent\textit{Proof: } It is clear that 
$(f_n,\phi^-_n)\subset\Gamma_{M,J}^k\times D^1$. Moreover, since 
$|\nabla\phi_n|^2=|\nabla\phi^-_n|^2+|\nabla\phi^+_n|^2$, we have 
$\mathcal{E}(f_n,\phi^-_n)\leq\mathcal{E}(f_n,\phi_n)$, whence 
$(f_n,\phi^-_n)$ is also a minimizing sequence. Now assume that 
(\ref{samelimit1}) is false. Then there exists a subsequence $\phi_n$ 
and $\lambda>0$ such that 
$\|\nabla\phi_{n}-\nabla\phi^-_{n}\|_{L^2}>\lambda$. Thus
\begin{eqnarray*}
\mathcal{E}(f_{n},\phi_{n})&=&E_{\mathrm{kin}}(f_{n},\phi_{n})+\frac{1}{2}\int_{\R^3}\left(|\nabla\phi^-_{n}|^2+|\nabla(\phi_{n}-\phi^-_{n})|^2\right)\,dx\\
&\geq&\mathcal{E}(f_{n},\phi^-_{n})+\frac{1}{2}\|\nabla\phi_{n}-\nabla\phi^-_{n}\|_{L^2}^2\geq\mathcal{E}(f_{n},\phi^-_{n})+\frac{\lambda^2}{2},
\end{eqnarray*}
which contradicts 
$\lim_{n\to\infty}\mathcal{E}(f_{n},\phi_{n})=\lim_{n\to\infty}\mathcal{E}(f_{n},\phi^-_{n})=I_{M,J}^k$.
\prfe

Let $(f_n,\phi_n)$ be a minimizing sequence. Since $(f_n,\phi_n)$ is 
bounded in $L^{1+1/k}\times D^1$, there exist $f_0\in L^{1+1/k}$, 
$\phi_0\in D^1$ and a subsequence $(f_n,\phi_n)$ such that
\[
f_n\rightharpoonup f_0\ \textnormal{in }L^{1+1/k},\quad 
\phi_n\rightharpoonup\phi_0\ \textnormal{in } L^6,\quad 
\nabla\phi_n\rightharpoonup\nabla\phi_0\ \textnormal{in } L^2
\]
and so, by \cite[Cor.~8.7]{LL},
\[
\phi_n\to\phi_0,\ \textnormal{pointwise almost everywhere.}
\]
By Lemma \ref{phinegative}, $\phi_0(x)\leq 0$, for almost all 
$x\in\R^3$ and by weak convergence, $f_0\geq0$ a.e.
It is clear that $(f_0,\phi_0)$ is our candidate for being a 
minimizer of the functional $\mathcal{E}$. In the next lemma we show 
that the energy functional is weakly lower semincontinuous..
\begin{lemma}\label{weaklowersemicon}
For all $0<k<2$, $J>0$ and $M>0$,
\[
I_{M,J}^k\geq\mathcal{E}(f_0,\phi_0).
\]
\end{lemma}
\noindent\textit{Proof: } Clearly,
\[
\liminf_{n\to\infty}\int_{\R^3}|\nabla\phi_n|^2dx\geq\int_{\R^3}|\nabla\phi_0|^2dx.
\]
Moreover
\[
\pzn\to\sqrt{e^{2\phi_0}+|p|^2},\quad\textnormal{pointwise a.e. (up 
to subsequences)}
\]
and so, for all $R>0$ and since $\phi_0$ is non-positive,
\[
\pzn f_n\rightharpoonup\sqrt{e^{2\phi_0}+|p|^2} f_0\ \textnormal{ in 
}L^{1+1/k}(B_R),
\]
where $B_R=\{(x,p)\in\R^6:|x|^2+|p|^2\leq R^2\}$. It follows that
\[
\int_{\R^3}\int_{B_R}\sqrt{e^{2\phi_0}+|p|^2}f_0\,dp\,dx\leq\liminf_{n\to\infty}\int_{\R^3} \int_{B_R}\sqrt{e^{2\phi_n}+|p|^2}f_n\,dp\,dx.
\]
We then have
\begin{eqnarray*}
\mathcal{E}(f_0,\phi_0)&=&\lim_{R\to\infty}\int_{\R^3} \int_{B_R}\sqrt{e^{2\phi_0}+|p|^2}f_0\,dp\,dx+\frac{1}{2}\int_{\R^3}|\nabla\phi_0|^2dx\\
&\leq&\lim_{n\to\infty}\mathcal{E}(f_n,\phi_n)=I_{M,J}^k,
\end{eqnarray*}
which completes the proof. 
\prfe

In the next lemma we provide a sufficient condition for the strong 
convergence of $f_n$ in $L^1$. 
\begin{lemma}\label{convergencef}
Let $(f_n,\phi_n)$ be a minimizing sequence and assume
\begin{equation}\label{assumption}
\|\nabla\phi_n-\nabla\phi_0\|_{L^2}\to 0,\ n\to\infty.
\end{equation}
Let $M>M_0$. Then
\[
\|f_n- f_0\|_{L^1}\to 0,\ \textnormal{ {\it as} }n\to\infty.
\]
Moreover, $(f_0,\phi_0)$ is a minimizer of the 
functional $\mathcal{E}$ over the space $\Gamma_{M,J}^k\times D^1$ and
\[
\|f_n- f_0\|_{L^{1+1/k}}\to 0,\ \textnormal{ {\it as} }n\to\infty.
\]
\end{lemma}
\noindent\textit{Proof: } Let $q=1+1/k>3/2$. To prove weak convergence 
in $L^1$ of $f_n$, it is clearly enough to show that no mass is lost at 
infinity, i.e., that for all $\varepsilon>0$, there exists 
$R(\varepsilon)>0$ such that
\[
\int_{\R^3}\int_{A_R}f_n\,dp\,dx\leq\varepsilon,
\]
where $A_R=\{(x,p)\in\R^6:\,|x|>R\textnormal{ or }|p|>R\}$. Since
\begin{eqnarray*}
\int_{\R^3}\int_{A_R}f_n\,dp\,dx&\leq& 
\int_{\R^3} \int_{|p|>R}f_n\,dp\,dx+\int_{|x|>R}\int_{\R^3} f_n\,dp\,dx\\
&\leq&\frac{1}{R}E_{\mathrm{kin}}(f_n,\phi_n)+\int_{|x|>R}\int_{\R^3} f_n\,dp\,dx
\end{eqnarray*}
and since the term $R^{-1}E_{\rm kin}$ can be made arbitrarily 
small---by taking $R$ sufficiently large---it is enough to prove that
\begin{equation}\label{nomassinf}
\forall\,\varepsilon>0,\ \exists 
R(\varepsilon)>0:\int_{|x|>R}\rho_{\!f_n}\,dp\,dx\leq\varepsilon. 
\end{equation}
If (\ref{nomassinf}) were false, then we could find $0<Q<M$, a 
subsequence $f_n$ and a sequence $R(n)$ (depending also on $Q$) such 
that
\[
\int_{|x|>R(n)}\int_{\R^3} f_n\,dp\,dx=Q,\quad\lim_{n\to\infty}R(n)=\infty.
\]
We write
\begin{eqnarray*}
Q&=&\int_{|x|>R(n)}\int_{\R^3} f_n\,dp\,dx\\ &\leqslant& \int_{|x|>R(n)}\int_{\R^3} e^{\phi_n}f_n\,dp\,dx-\int_{|x|>R(n)}\int_{\R^3} \phi_nf_n\,dp\,dx\\
&\leqslant& \int_{|x|>R(n)}\int_{\R^3}\pzn 
f_n\,dp\,dx+ \int_{|x|>R(n)}|\phi_n|\int_{\R^3}f_n\,dp\,dx,
\end{eqnarray*}
where the simple bound $\exp(\phi_n)-\phi_n\geqslant 1$ has been 
used. By (\ref{sobolev}) and (\ref{assumption}), 
$\|\phi_n-\phi_0\|_{L^6}\to 0$, $n\to\infty$; whence, for all 
$\varepsilon>0$ and sufficiently large $n$,
\[
\int_{|x|>R(n)}|\phi_n|^6\,dx\leqslant\varepsilon.
\]
Thus, by Lemma \ref{estrho} and H\"older's inequality,
\[
\int_{|x|>R(n)}|\phi_n|\int_{\R^3} f_n\,dp\,dx\leqslant\varepsilon(n),
\]
where $\varepsilon(n)\to 0$, as $n\to\infty$. In conclusion
\[
Q\leqslant \int_{|x|>R(n)}\int_{\R^3} \pzn f_n\,dp\,dx+\varepsilon(n).
\]
We shall prove that the above inequality leads to the contradiction 
that $I_{M,J}^k$ is not the energy infimum in the space 
$\Gamma_{M,J}^k$. To this purpose, we define the sequence
\[
\widetilde{f}_n=\alpha f_n(\beta x, p)\,\chi_{\{|x|\leqslant \beta^{-1}R(n)\}},
\qquad
\widetilde{\phi}_n = \phi_n(\beta x),
\]
where
\[
\alpha=\beta^2,\ \beta=\frac{M-Q}{M}<1.
\]
It is easy to check that the sequence $(\widetilde{f}_n,\widetilde{\phi}_n)$ is contained in 
$\Gamma_{M,J}^k\times D^1$.
Moreover
\begin{eqnarray*}
\mathcal{E}(\widetilde{f}_n,\widetilde{\phi}_n)&=& 
\beta^{-1}\left( \int_{|x|\leqslant 
R(n)}\int_{\R^3}\sqrt{e^{2\phi_n}+|p|^2}f_n\,dp\,dx+\frac{1}{2}\int_{\R^3}|\nabla\phi_n|^2\right)\\
&=& 
\beta^{-1}\left(\mathcal{E}(f_n,\phi_n)-\int_{|x|>R(n)}\int_{\R^3} \pzn 
f_n\,dp\,dx\right)\\
&\leqslant&\left(\frac{M}{M-Q}\right) 
\left(\mathcal{E}(f_n,\phi_n)-Q+\varepsilon(n)\right).
\end{eqnarray*}
Letting $n\to\infty$ we obtain
\[
\liminf_{n\to\infty}\,\mathcal{E}(\widetilde{f}_n,\widetilde{\phi}_n)\leqslant\left(\frac{M}{M-Q}\right)
\left(I_{M,J}^k-Q\right).
\]
Let us denote by $F(Q)$ the function in the right hand side of the 
latter inequality. It satisfies $\lim_{Q\to 0^+}F(Q)=I_{M,J}^k$, 
$\lim_{Q\to M^-}F(Q)=-\infty$ and
\[
F'(Q)=\left(\frac{M}{M-Q}\right)  \left(\frac{I_{M,J}^k-M}{M-Q}\right)<0,\
\]
for all $Q\in (0,M)$ (because of the assumption  $M>M_0$), which 
leads to the contradiction 
$\lim_{n\to\infty}\mathcal{E}(\widetilde{f}_n,\widetilde{\phi}_n)<I_{M,J}^k$. 
This concludes the proof that $f_n\rightharpoonup f_0$ in $L^1$. By 
weak convergence, $\|f_0\|_{L^1}=M$ and $\|f_0\|_{L^q}\leq J$. Thus $(f_0,\phi_0)$ is a minimizer by Lemma \ref{weaklowersemicon}. By Proposition \ref{boundminimizer}, 
$\|f_0\|_{L^{1+1/k}}=J$. Hence 
$\liminf_{n\to\infty}\|f_n\|_{L^{1+1/k}}=\|f_0\|_{L^{1+1/k}}$ and so, after 
possibly extracting a subsequence, $f_n\to f$, strongly in 
$L^{1+1/k}$ (see for instance \cite[Thm.~2.11]{LL}). 
Extracting a subsequence, the convergence also holds pointwise almost 
everywhere. A standard application of Egoroff's theorem shows that 
weak convergence in $L^1$ and almost everywhere pointwise convergence 
of a sequence of functions imply strong convergence in $L^1$. This 
concludes the proof of the proposition.
\prfe

By virtue of Lemma 
\ref{convergencef}, Theorem \ref{existence} will follow if we prove 
that minimizing sequences, after properly translated in space, satisfy (\ref{assumption}). Our strategy is to 
show that it suffices to prove this for a special class of minimizing 
sequences, which enjoy some additional regularity and to which one can apply arguments similar to those valid for the Vlasov-Poisson system.

\begin{lemma}\label{minphi}
Let $f\in\Gamma_{M,J}^k$ being fixed such that $E_{\rm kin}(f,0)<\infty$.
\begin{itemize}
\item[(i)] The variational problem
\[
\mathcal{I}_{\! f}=\inf_{\psi\in D^1}\mathcal{J}_{\!
f}(\psi),\quad\mathcal{J}_{\! f}(\psi)=\mathcal{E}(f,\psi),
\]
has a unique minimizer $\psi_{\! f}\in D^1$. Moreover, $\psi_{\!f}$ is a non-positive continuous function and satisfies
\begin{equation}\label{poisson}
\bigtriangleup\psi=e^{2\psi}\int_{\R^3}\frac{f}{\sqrt{e^{2\psi}+|p|^2}}\,dp,
\end{equation}
in the sense of distributions.
\item[(ii)] For all $\phi\in D^1$ we have
\[
\mathcal{J}_{\! f}(\phi)-\mathcal{I}_{\! 
f}\geq\frac{1}{2}\|\nabla\phi-\nabla\psi_{\! f}\|_{L^2}^2.
\]
\end{itemize}
\end{lemma}
\noindent\textit{Proof: } Let $\psi_n\in D^1$ be a minimizing sequence 
for the functional $\mathcal{J}_{\! f}$; there exists $\psi_{\!
f}\in D^1$ and a subsequence---still denoted $\psi_n$---such that 
$\psi_n\rightharpoonup\psi_{\! f}$ in $L^6$, 
$\nabla\psi_n\rightharpoonup\nabla\psi_{\! f}$ in $L^2$ and 
$\psi_n\to\psi_{\! f}$ pointwise almost everywhere. Moreover, by 
Lemma \ref{phinegative} (with $f_n$ kept fixed), $\psi_{\!
f}\leq 0$ a.e. We prove that $\psi_{\! f}$ is a minimizer by 
showing that the functional $\mathcal{J}_{\! f}$ is weakly lower 
semicontinuous. Clearly
\[
\liminf_{n\to\infty}\int_{\R^3}|\nabla\psi_n|^2dx\geq\int_{\R^3}|\nabla\psi_{\!
f}|^2dx.
\]
Moreover, by convexity of the function $x\to\sqrt{e^{2x}+|p|^2}$,
\[
E_{\rm kin}(\psi_{\! f},f)-E_{\rm kin}(\psi_n,f)\leq \int_{\R^3} 
(\psi_{\!f}-\psi_n)\int_{\R^3}\frac{e^{2\psi_{\! 
f}}}{\sqrt{e^{2\psi_{\! f}}+|p|^2}}f\,dp\,dx.
\]
The function $e^{2\psi_{\! f}}\int_{\R^3}(e^{2\psi_{\!
f}}+|p|^2)^{-1/2}f\,dp$ is dominated by $\rho_{\! f}$ and so it 
belongs to $L^{6/5}$ by Lemma \ref{estrho}. By weak convergence, the 
right hand side of the last inequality converges to zero as 
$n\to\infty$, whence
\[
\liminf_{n\to\infty}E_{\rm kin}(f,\psi_n)\geq E_{\rm kin}(f,\psi_{\! f})
\]
and the proof that $\psi_{\! f}$ is a minimizer is complete. 
Moreover, $\psi_{\! f}$ is a weak solution of the Euler-Lagrange 
equation for the functional $\mathcal{J}_{\! f}$, which is 
(\ref{poisson}). By Lemma \ref{estrho}, the right hand side of 
(\ref{poisson}) is in $L^{j}$, where $j>3/2$ is given by (\ref{j}). 
Whence $\psi_{\! f}\in W^{2,j}_{loc}(\R^3)$, which is continuously 
embedded in $C(\R^3)$. In particular, $\psi_{\! f}$ is bounded. 
Since the functional $\mathcal{J}_{\! f}$ is uniformly convex in 
$D^1\cap L^\infty$, uniqueness of minimizers follows by standard 
theory of calculus of variations (see \cite{E}, for instance). It 
remains to prove (ii). For this purpose we write
\begin{eqnarray*}
\mathcal{J}_{\! f}(\phi)-\mathcal{I}_{\! f}&=&\mathcal{J}_{\!
f}(\phi)-\mathcal{J}_{\! f}(\psi_{\!
f})\geq\int_{\R^3}\left(\phi-\psi_{\!
f}\right)\int_{\R^3}\frac{e^{2\psi_{\! f}}}{\sqrt{e^{2\psi_{\!
f}}+|p|^2}}\,f\,dp\,dx\\
&&+\int_{\R^3}\nabla\psi_{\! f}\cdot\left(\nabla\phi-\nabla\psi_{\!
f}\right)\,dx+\frac{1}{2}\int_{\R^3}|\nabla\phi-\nabla\psi_{\! f}|^2dx\\
&=&\frac{1}{2}\|\nabla\phi-\nabla\psi_{\! f}\|_{L^2}^2,
\end{eqnarray*}
where the convexity of $x\to\sqrt{e^{2x}+|p|^2}$ and the 
Euler-Lagrange equation (\ref{poisson}) have been used. This 
completes the proof of the lemma.
\prfe

\begin{lemma}\label{poissonequation}
Let $f\in\Gamma_{M,J}^k$ being fixed such that $E_{\rm 
kin}(f,0)<\infty$. The equation (\ref{poisson}) has a unique solution 
$\psi_{\! f}\in D^1$. Moreover, $\psi_{\! f}$ is a non-positive 
continuous function and satisfies
\[
\psi_{\! f}(x)=-\int_{\R^3}\frac{e^{2\psi_{\!
f}(y)}}{|x-y|}\int_{\R^3}\frac{f(y,p)}{\sqrt{e^{2\psi_{\! 
f}(y)}+|p|^2}}\,dp\,dy,
\]
\[
\nabla\psi_{\! f}=\int_{\R^3}\frac{(x-y)}{|x-y|^3}e^{2\psi_{\! 
f}(y)}\!\!\int_{\R^3}\frac{f(y,p)}{\sqrt{e^{2\psi_{\!
f}(y)}+|p|^2}}\,dp\,dy,
\]
\begin{eqnarray*}
\int_{\R^3}|\nabla\psi_{\! f}|^2dx&=&\int_{\R^3} \frac{e^{2(\psi_{\!
f}(x)+\psi_{\!
f}(y))}}{|x-y|}\left(\int_{\R^3}\frac{f(x,p)}{\sqrt{e^{2\psi_{\! 
f}(x)}+|p|^2}}\,dp\right)\\
&&\times\left(\int_{\R^3}\frac{f(y,p')}{\sqrt{e^{2\psi_{\!
f}(y)}+|p'|^2}}\,dp'\right)\,dx\,dy.
\end{eqnarray*}
\end{lemma}
\noindent\textit{Proof: } The existence claim follows directly from 
Lemma \ref{minphi}. To prove uniqueness, it suffices to show that any 
solution of (\ref{poisson}) is a minimizer of the functional 
$\mathcal{J}_{\! f}$. This follows by the inequality
\[
\mathcal{I}_{\! f}-\mathcal{J}_{\! 
f}(\psi)\geq\frac{1}{2}\|\nabla\psi-\nabla\psi_{\! f}\|_{L^2}^2,
\]
which is valid for all solutions $\psi\in D^1$ of (\ref{poisson}) and 
which is derived by using the convexity of $x\to\sqrt{e^{2x}+|p|^2}$. 
Now set $\widetilde{f}(x,p)=e^{4\psi}f(x,e^{\psi}p)$. So doing, 
(\ref{poisson}) becomes the {\it linear} Poisson equation 
$\bigtriangleup\psi=\int_{\R^3}\widetilde{f}(1+|p|^2)^{-1/2}dp$. Note that 
$\widetilde{f}$ has the same $L^q$ regularity of $f$. Moreover 
$\int_{\R^3}\widetilde{f}|p|\,dp=\int_{\R^3} f|p|\,dp<\infty$, whence, by Lemma 
\ref{estrho}, $\int_{\R^3}\widetilde{f}(1+|p|^2)^{-1/2}dp\in L^j$. The rest 
of the lemma follows by standard potential theory, see 
for instance \cite[Thm.~6.21]{LL}. 
\prfe

\begin{corollary}\label{coro}
Let $(f_n,\phi_n)\subset\Gamma_{M,J}^k\times D^1$ be a minimizing 
sequence for the functional $\mathcal{E}$ and let $\psi_{\! f_n}\in 
D^1$ be the unique weak solution of
\begin{equation}\label{poissonn}
\bigtriangleup\psi=e^{2\psi}\int_{\R^3}\frac{f_n}{\sqrt{e^{2\psi}+|p|^2}}\,dp.
\end{equation}
Then $(f_n,\psi_{\! f_n})$ is still a minimizing sequence and
\begin{equation}\label{psiphi}
\|\nabla\psi_{\! f_n}-\nabla\phi_n\|_{L^2}\to 0,\ n\to\infty.
\end{equation}
In particular, $\psi_{\! f_n}\to\phi_0$, as $n\to\infty$, pointwise 
almost everywhere.
\end{corollary}
\noindent\textit{Proof: } As 
$\mathcal{E}(f_n,\phi_n)\geq\inf\{\mathcal{J}_{\!
f_n}(\psi),\,\psi\in D^1\}=\mathcal{E}(f_n,\psi_{\! f_n})$, then 
$(f_n,\psi_{\! f_n})$ is still a minimizing sequence. Moreover, by 
Proposition \ref{minphi} (ii) we have
\[
\mathcal{E}(f_n,\phi_n)-\mathcal{E}(f_n,\psi_{\! f_n})\geq
\frac{1}{2}\|\nabla\phi_n-\nabla\psi_{\! f_n}\|_{L^2}^2
\]
and since the left hand side converges to zero, (\ref{psiphi}) is 
proved. Thus, after possibly extracting a subsequence, $(\psi_{\! 
f_n}-\phi_n)\to 0$, $n\to\infty$, pointwise a.e. and since 
$\phi_n\to\phi_0$ in the same sense, the proof of the corollary is 
complete.
\prfe

Thus the problem of proving strong convergence for $\nabla\phi_n$ in $L^2$ up to spatial translations has now been reduced to that of proving the same claim for the sequence $\psi_{\!f_n}$. This problem will be addressed next, using the deep result by Burchard and Guo \cite{BG}. 

\begin{lemma}\label{hypBG} Let $(f_n, \psi_{\! f_n})$ be a minimizing sequence to the functional $\mathcal{E}$ given by Corollary \ref{coro}. Then, there exits $\phi_0$ in $D^1$ such that
\begin{itemize}
\item[i)] $\lim_{n \to \infty} \| \nabla  \psi_{\! f_n}^* - \nabla \phi_0 \|_{L^2}^2 = 0 $,
\item[ii)] $\lim_{n \to \infty}  \| \nabla  \psi_{\! f_n} \|_{L^2}^2  = \| \nabla \phi_0 \|_{L^2}^2 $.
\end{itemize}
\end{lemma}
\noindent\textit{Proof: } Thanks to Lemma \ref{boundminimizer} we deduce that $(f_n^*, -\psi_{\! f_n}^*)$ is also a minimizing sequence and verifies the inequalities (\ref{41}) and (\ref{42}). By uniqueness of solutions to (\ref{poisson}), see Lemma \ref{poissonequation}, $\psi_{\!f^*_n}$ is also spherically symmetric. Moreover, $\psi_{\!f^*_n}=-\psi_{\! f_n}^*$, by uniqueness of minimizers to the variational problem in Lemma \ref{minphi}. 
We now prove the first assertion of the Lemma. We notice first that since $\nabla\psi_{\!f_n}\in W^{1,j}(B_R)$, for any ball $B_R=\{x\in\R^3:|x|\leq R\}$ and the inclusion $W^{1,j}(B_R)\subset L^2(B_R)$ is compact, then $\nabla\psi_{\!f_n}$ converges strongly in $L^2(B_R)$ (up to subsequences). Thus the only property that we need to prove is that
\begin{equation}\label{property}
\int_{|x|>R}|\nabla\psi_{n}^*|\to 0\quad\textnormal{as }R\to\infty,
\end{equation} 
where we denote $\psi_{\! f_n}^*=\psi_n^*$. Integrating (\ref{poisson}) we obtain 
\[
{\psi_{n}^*}'(r)=\frac{1}{r^2}\int_0^rs^2e^{2\psi^*_{n}(s)}\int_{\R^3}\frac{f^*_n}{\sqrt{e^{2\psi^*_{n}(s)}+|p|^2}}\,dp\,ds,
\]
from which it follows that
\[
|{\psi^*_{n}}'(r)|\leq \frac{M}{r^2}
\]
and the property (\ref{property}) is then satisfied by the sequence $\psi_n^*$. In conclusion, the sequence $\nabla\psi^*_n$ converges strongly in $L^2$. 

The second assertion can be easily deduced from the above arguments and using (\ref{41}) and (\ref{42}) .
\prfe

Under the framework of Lemma \ref{hypBG}, the hypotheses of Theorem 2 in \cite{BG} hold and as a consequence there is a sequence of spatial translations $T_n$ such that
\[
\lim_{n \to \infty} \| T_n\nabla\psi_{\!f_n}  - \nabla \phi_0 \|_{L^2}^2 = 0. 
\]
This result coincides precisely with the hypothesis of Lemma \ref{convergencef} and, hence, allows to  conclude the proof of Theorem \ref{existence}. 
\section{Properties of minimizers}\label{properties}
The aim of this section is to prove Theorem \ref{Lagrange}. We start by showing that the minimizers constructed in the previous section are isotropic polytropes solutions of the Nordstr\"om-Vlasov system. For this purpose we use the  method of the Lagrange multipliers. Note that we are combining some previous established techniques, see for example \cite{GR1,GR2,R2,R3}, together with precise change of scales that preserves both constraints in our minimizing problem.

Let $(f_0,\phi_0)\in\Gamma_{M,J}^k\times D^1$ be a minimizer and for any $\varepsilon>0$ fixed define $S_\varepsilon=\{(x,p)\in\R^6:\varepsilon\leq f_0(x,p)\leq \varepsilon^{-1}\}$. Let $\eta\in L^\infty(\R^6)$ be a real valued function with compact support such that $\eta\geq0,\textnormal{ a.e. for }(x,p)\in\R^6\setminus\textrm{supp}f_0$ and $\textrm{supp}\,\eta\subseteq\left(\R^6\setminus\textrm{supp}f_0\right)\cup S_\varepsilon$. For $t\in[0,T]$ and $T=(\|\eta\|_1+\|\eta\|_q+\|\eta\|_\infty)^{-1}\varepsilon/2$ we define
\[
f_t(x,p)=\alpha(t)^{3}M\frac{f_0+t\eta}{\|f_0+t\eta\|_1}(\alpha(t) x,p),\quad \phi_t(x)=\phi_0(\alpha(t) x),
\]
where 
\[
\alpha(t)=\left(\frac{J}{M}\frac{\|f_0+t\eta\|_1}{\|f_0+t\eta\|_q}\right)^{\frac{q}{3q-3}},\quad q=1+1/k>3/2.
\]
Note that $f_0+t\eta\geq0$ a.e. and, for a $\varepsilon$ small enough, 
\[
M/2\leq\|f_0+t\eta\|_1\leq M+\varepsilon/2,\quad J/2\leq\|f_0+t\eta\|_q\leq J+\varepsilon/2. 
\] 
{}From this we infer that $\alpha$ is a smooth function on $[0,T]$ and 
\[
\alpha'(t)=\frac{q}{3q-3}\alpha(t)\left[\frac{\int_{\R^3}\int_{\R^3}\eta\,dp\,dx}{\|f_0+t\eta\|_1}-\frac{\int_{\R^3}\int_{\R^3}(f_0+t\eta)^{q-1}\eta\,dp\,dx}{\|f_0+t\eta\|_q^q}\right].
\]
Moreover $\sup_{[0,T]}\alpha''(t)$ is bounded.
By inspection, $(f_t,\phi_t)\in\Gamma_{M,J}^k\times D^1$, for all $t\in [0,T]$ and
\begin{eqnarray}
\mathcal{E}(f_t,\phi_t)-\mathcal{E}(f_0,\phi_0)&=&\left(\frac{M}{\|f_0+t\eta\|_1}-1\right)E_{\rm kin}(f_0,\phi_0)\label{temporale}\\
&&\hspace{-1cm}+\frac{Mt}{\|f_0+t\eta\|_1}E_{\rm kin}(\eta,\phi_0)+\frac{1}{2}\left(\frac{1}{\alpha(t)}-1\right)\int_{\R^3}|\nabla\phi_0|^2dx.\nonumber
\end{eqnarray}
By a Taylor expansion at $t=0^+$ and straightforward estimates we obtain
\begin{eqnarray*}
&&\frac{M}{\|f_0+t\eta\|_1}-1=-\frac{t}{M}\int_{\R^3}\int_{\R^3}\eta\,dp\,dx+O(t^2),\quad \frac{Mt}{\|f_0+t\eta\|_1}=t+O(t^2),\\
&&\frac{1}{\alpha(t)}-1=-t \frac{q}{3q-3}\left[\frac{\int_{\R^3}\int_{\R^3}\eta}{M}-\frac{\int_{\R^3}\int_{\R^3}f_0^{q-1}\eta}{J^q}\right]+O(t^2),
\end{eqnarray*}
where the notation $O(t^2)$, as usual, means that the rest terms are bounded by $Ct^2$, for a positive constant $C$ depending on $\varepsilon,\,f_0$ and $\eta$, but not on $t$. Substituting into (\ref{temporale}) we get
\[
\mathcal{E}(f_t,\phi_t)-\mathcal{E}(f_0,\phi_0)=t\int_{\R^3}\int_{\R^3}\left(E-E_0+c f_0^{q-1}\right)\eta\,dp\,dx+O(t^2),
\]
where $E_0$ and $c$ are given  in Theorem \ref{Lagrange}. Recalling the class of admissible test functions $\eta$ and the fact that $\varepsilon>0$ is arbitrary, we conclude that $E-E_0\geq0$ a.e. on $\R^6\setminus\textrm{supp}f_0$ and $f_0=(\frac{E_0-E}{c})^{1/(q-1)}$ a.e. on $\textrm{supp}f_0$, which proves the first part of Theorem \ref{Lagrange}. As to the admissible range for $E_0$ claimed in the theorem, we notice that, by the Virial Theorem and the definition of $E_0$,
\begin{eqnarray*}
\int_{\R^3}  \int_{\R^3} e^{2 \phi_0} \frac{f_0}{E} \,  dp \,dx&=&
\left (I_{M,J}^k - \int_{\R^3} |\nabla \phi_0|^2 \, dx\right)
\\ &=&\left (\frac{6}{2-k}E_0 M -  \frac{k+4}{2-k} I_{M,J}^k \right).
\end{eqnarray*}
The above quantity  and the constant $c$ are both positive, { from which
the admissible range of $E_0$ is derived.}

We proceed proving some properties of the minimizers. By Lemma \ref{poissonequation},
\[
|\phi_0(x)|\leq\int_{\R^3} e^{2\phi_0(y)}\int_{\R^3}\frac{f_0(y,p)}{\sqrt{e^{2\phi_0(y)}+|p|^2}}\,dp\,\frac{dy}{|x-y|}.
\]
For all $R>1$ we split the integral in the right hand side according to $|x-y|\leq 1/R$, $1/R\leq |x-y|\leq R$, $|x-y|\geq R$. Straightforward estimates lead to
\[
|\phi_0(x)|\leq C\|\mu_{\! f_0}\|_jR^{\frac{3-7j}{j}}+R\int_{|y|\geq|x|-R}\rho_{\! f_0}(y)\,dy+\frac{M}{R},
\] 
which yields $\lim_{|x|\to\infty}\phi_0(x)=0$. From this and the fact that $E_0<1$, it follows that $\pzo>E_0$, for $|p|>1$ or for $|x|$ large enough, which proves the compact support property of $f_0$. Using $f_0=((E-E_0)/c)_+^k$ and a change of variable entail
\[
e^{2\phi_0}\int_{\R^3}\frac{f_0}{\pzo}\,dp=4\pi c^{-k}\int_{e^{\phi_0}}^{E_0}\sqrt{E^2-e^{2\phi}}(E_0-E)^kdE\leq C.
\]
This result together with Lemma \ref{poissonequation} implies that $\phi_0\in W^{2,\infty}_{loc}$, whence the characteristics of the time independent Vlasov equation are well defined $C^1$ curves. Since $f_0$ depends only on the particles energy $E=\pzo$, then it is constant along characteristics, which is the definition of mild solution. 
Next we show that $(f_0,\phi_0)$ is spherically symmetric with respect to some point in $\R^3$. By using Lemma \ref{boundminimizer}, $(f_0^*,-\phi_0^*)$ is also a minimizer. Then 
\[
\int_{\R^3}|\nabla\phi_0^*|^2dx=\int_{\R^3}|\nabla\phi_0|^2dx
\]
and $(f_0^*,-\phi_0^*)$ is also a solution of (\ref{Poisson}). Writing $\phi_0^*(x)=\psi(r)$, $r=|x|$, and integrating (\ref{Poisson}) we obtain
\[
\psi'(r)=\frac{1}{r^2}\int_0^rs^2e^{2\psi(s)}\int_{\R^3}\frac{f^*_0}{\sqrt{e^{2\psi(s)}+|p|^2}}\,dp\,ds,
\]
where $\psi'=d\psi/dr$.
From the previous equation we deduce that $\nabla\phi_0^*=(x/r)\psi'\neq 0$ a.e., whence $\phi_0(x)$ coincides with a spatial translation of $\phi_0^*(x)$, see \cite{BZ}. Since $E_0$ is the same for $f_0$ and $f_0^*$, we conclude by using (\ref{polytropes}) that $f_0$ equals $f_0^*$, up to a translation in space.

To conclude the proof of Theorem \ref{Lagrange}, it remains to establish the uniqueness statement. Without loss of generality, we assume that the potential function is spherically symmetric with respect to the origin, i.e., $\psi(r) = \phi_0(x), \ r = |x|$; recall that uniqueness is up to a spatial translation. The non-linear Poisson equation (\ref{Poisson}) for $\psi(r)$ is
\begin{equation}
(r^2 \psi'(r))' =   r^2 e^{2 \psi(r)} \int_{\R^3} \frac{f_0}{E} \,dp.
 \label{poisonrad}
\end{equation}  
On the other hand, $\psi$ is strictly increasing and vanishes at infinity  which assures the existence of 
$r_0 \in \mathbb{R}_+$ such that  
\begin{equation}
\psi(r_0) = \log{E_0}.
\label{psir0}
\end{equation} 
Let us note that if $\psi(0) > \log{E_0}$, then $\psi$ is constant in time 
since $f_0 $ vanishes  when $e^{ \psi(r)} > E_0$.
Now, we prove that $r_0$, $\psi(r_0)$ and $\psi'(r_0)$ are uniquely determined
by $E_0$ for any minimizer. Let $r\geq r_0$;
using the Virial Theorem and the definition of $E_0$ we find
\begin{eqnarray}
\int_0^r s^2 e^{2 \psi(s)} \int_{\R^3} \frac{f_0}{E} \, dp \,ds
& = &
\int_0^\infty s^2 e^{2 \psi(s)} \int_{\R^3} \frac{f_0}{E}\, dp \, ds 
\nonumber \\
&
= 
& 
\frac{1}{4\pi} \int_{\R^3}  \int_{\R^3} e^{2 \phi_0(x)} \frac{f_0}{E} \,  dp \,dx
\nonumber \\
& = &
\frac{1}{4\pi} \left (\frac{6}{2-k}E_0 M -  \frac{k+4}{2-k} I_{M,J}^k \right),
\quad \forall \ r \geq r_0.
\nonumber
\end{eqnarray}
Then, integrating (\ref{poisonrad}) we have
\begin{equation}
\psi'(r) 
= 
\frac{1}{4\pi\,r^2} \left (\frac{6}{2-k}E_0 M -  \frac{k+4}{2-k} I_{M,J}^k \right),
\quad \forall \ r \geq r_0\, .
\label{psiprima}
\end{equation}
Using  this expression we obtain 
$$\int_{r_0}^\infty 4 \pi s^2 (\psi'(s) )^2 ds 
=
\frac{1}{4\pi} \left (\frac{6}{2-k}E_0 M -  \frac{k+4}{2-k} I_{M,J}^k \right)^2 \frac{1}{r_0}
$$
or 
$$\int_{r_0}^\infty 4 \pi s^2 (\psi'(s) )^2 ds 
 =  - \psi(r_0)   \int_0^{r_0} \int_{\R^3} e^{2 \psi(\xi)} \frac{f_0}{E} \,  dp \,d\xi
$$
which allow to prove that 
\begin{equation}
r_0 = \frac{-1}{4 \pi \log{E_0}} \left (\frac{6}{2-k}E_0 M -  \frac{k+4}{2-k} I_{M,J}^k \right)\, 
\label{r0}
\end{equation}
and 
\begin{equation}
\psi'(r_0)  
= 4 \pi (\log{E_0})^2  \left (\frac{6}{2-k}E_0 M -  \frac{k+4}{2-k} I_{M,J}^k \right)^{-1}\, .
\label{psiprimar0}
\end{equation}
The existence and uniqueness theory
developed in  \cite{C} for (\ref{poisonrad})  implies that any minimizer  can be seen  
as the unique solution to (\ref{poisonrad}) with initial conditions $\psi(r_0)$ 
and $\psi'(r_0)$. Since these quantities are uniquely parametrized by $E_0$
(see  (\ref{psir0}), (\ref{r0}) and (\ref{psiprimar0}))
the question about the uniqueness of the minimizer  
(up to spatial translations) is now equivalent to prove that the spherically symmetric minimizer is determined by an unique $E_0$.

For any $E_0 \in (\frac{k+4}{6} \frac{I_{M,J}^k}{M},\frac{I_{M,J}^k}{M})$ we 
have a solution $\psi = \psi_{E_0} $ to (\ref{poisonrad}) with initial conditions 
given by (\ref{psir0}), (\ref{r0}) and (\ref{psiprimar0}).
Associated to every $\psi_{E_0}$  there is a density function $f = f_{E_0}$  defined by  
(\ref{polytropes}).
At this point we  check  that the mass associated with
the functions $f_{E_0}$ is monotone as function of $E_0$,
which entails uniqueness of the minimizer by the 
condition $\|f_0\|_1 =M$. 
Unfortunately, these quantities can not be computed directly, since the initial 
conditions depends on several parameters, $M$, $J$, and $k$ and 
unknown quantities like $I_{M,J}^{k}$. 

In order to skip the  explicit dependence of equation (\ref{poisonrad})  on $E_0$ 
{ we scale the potential as}
\begin{equation} \label{scaling2}
\widetilde{\psi}(r)  = \psi(b r) - \log E_0  , \  \mbox{ where }  \  b= \frac{c^{k/2}}{E_0^{2 + k/2}}  .
\end{equation}
Now $\widetilde{\psi}(r)$ verifies
\begin{equation}
(r^2 \widetilde \psi'(r))' =   r^2 e^{2 \widetilde \psi(r)} \int_{\R^3} \frac{\widetilde{f}}{\widetilde{E}} \,dp   ,
 \label{poisonradscaled}
\end{equation}  
with 
$\widetilde{E}= \sqrt{e^{2 \widetilde \psi(r)} + |p|^2}$,  $\widetilde{f} = (1 - \widetilde{E})^k_+$
and initial conditions
\begin{equation}
\widetilde{\psi}(\tilde{r}_0) = 0\, , \quad
\widetilde{\psi}'(\tilde{r}_0) = b \psi'(r_0)\, , \quad
\tilde{r}_0 = r_0 /b \,.
\label{icscaled} 
\end{equation}

\begin{figure}
\hspace{-0.68cm}\resizebox{0.65\hsize}{!}{\includegraphics*{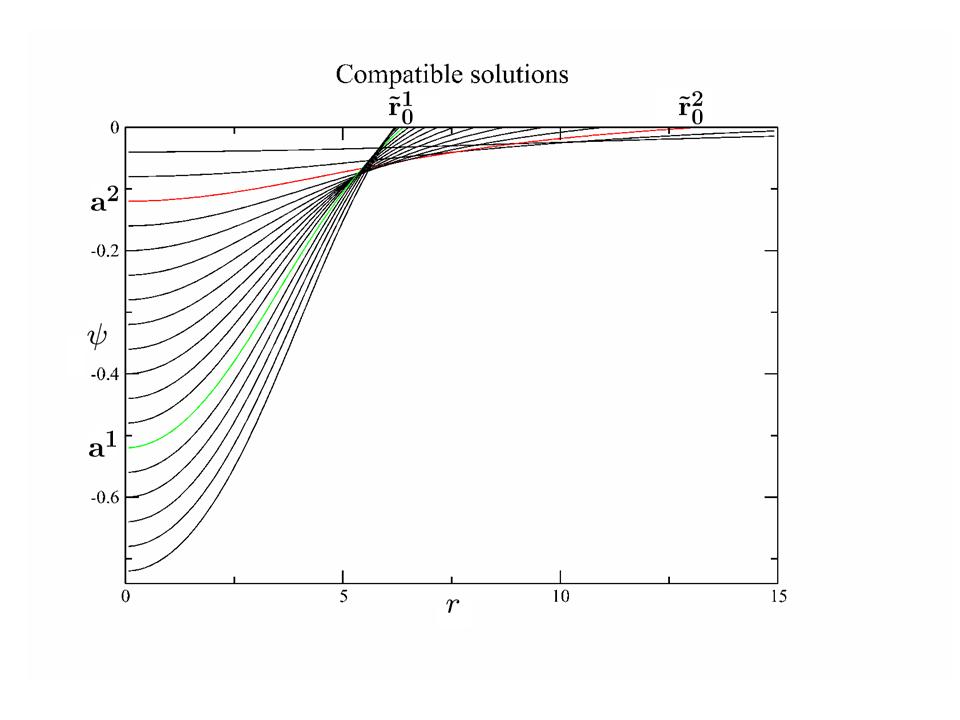}}
\hspace{-1.5cm}
\resizebox{0.65\hsize}{!}{\includegraphics*{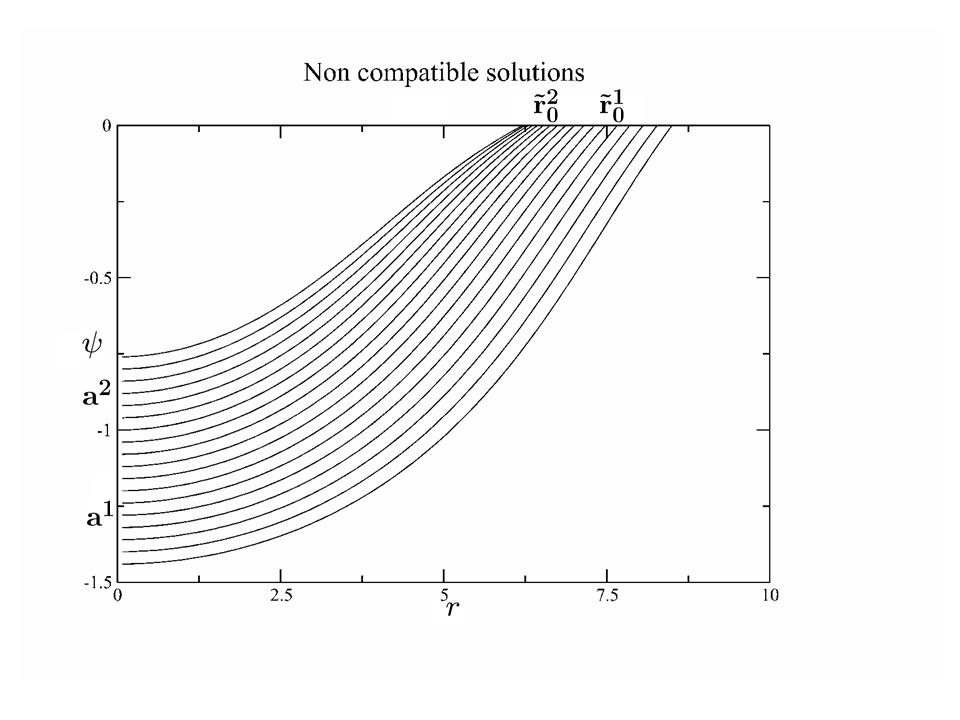}}
\caption{Families of solutions to  (\ref{poisonradscaled})}
\label{fig1}
\end{figure}
Since the initial conditions  (\ref{icscaled})  cannot be explicitly calculated, we consider numerical  simulations with initial conditions at $r=0$,
\begin{equation}
\widetilde{\psi}(0) = \psi(0) -\ln E_0 := a  \leq 0\, , \quad
\widetilde{\psi}'(0) =b \psi'(0) = 0
\label{ic0scaled} 
\end{equation}
parametrized in terms of the new parameter $a$ whose relation with $E_0$ is at this moment unknown. Figure \ref{fig1} shows these solutions for some values of $a$ in the case $k=1$. 
Two different regimes have been detected numerically for several values of $k$. 
For small values of $a$ ($a \leq a^* \approx 0.723$ when $k=1$) 
the solutions intersect themselves; in particular, for initial conditions 
$a^1 < a^2 (< a^*)$, the correspoding solutions $\widetilde{\psi}_1$ and $\widetilde{\psi}_2$ satisfy
\[
\tilde{r}_0^1< \tilde{r}_0^2,\quad\quad \widetilde{\psi}_1' (\tilde{r}_0^1) > \widetilde{\psi}_2' (\tilde{r}_0^2),
\]
where $\tilde{r}_0^1$, $\tilde{r}_0^2$ are the points such that
$\widetilde{\psi}_1 (\tilde{r}_0^1)= \widetilde{\psi}_2 (\tilde{r}_0^2) = 0$.
However, for large values of $a$ ($a \geq 0.723$ when $k=1$) the solutions mantain 
the order meanwhile they reach negative values.
In particular for initial conditions  $a^1 < a^2$ then 
$\widetilde{\psi}_1 (r) < \widetilde{\psi}_2 (r) $ for $r < \tilde{r}_0^1$
and
\[
\tilde{r}_0^2< \tilde{r}_0^1, \quad\quad\widetilde{\psi}'_2 (\tilde{r}_0^2) < \widetilde{\psi}'_1 (\tilde{r}_0^1)
\]
holds. It is not hard to see that the second type of solutions
is not compatible with conditions  (\ref{icscaled}). In fact, by the relations (\ref{r0}), (\ref{psiprimar0}) and (\ref{icscaled}) one can easily deduce that $\widetilde{\psi}'(\tilde{r}_0)$ must be a decreasing function of $\tilde{r}_0$.
The solutions with small values of $a$ 
corresponding to conditions (\ref{icscaled}) are thus the correct ones. We infer that the initial conditions in $r=0$ verify a strictly increasing relation with the 
values of $E_0$, that is, 
$$E_0^1 < E_0^2 \ \in   \left(\frac{k+4}{6} \frac{I_{M,J}^k}{M},\frac{I_{M,J}^k}{M} \right)
\quad \Longrightarrow \quad
\widetilde{\psi}_1(0) < \widetilde{\psi}_2(0)
$$
where  $\widetilde{\psi}_1$ and  $\widetilde{\psi}_2$ are the solutions to 
(\ref{poisonradscaled}) with initial conditions (\ref{icscaled}) (determined by $E_0^1$  and 
$E_0^2$). 

\begin{figure}
\hspace{2.2cm}\resizebox{0.7\hsize}{!}{\includegraphics*{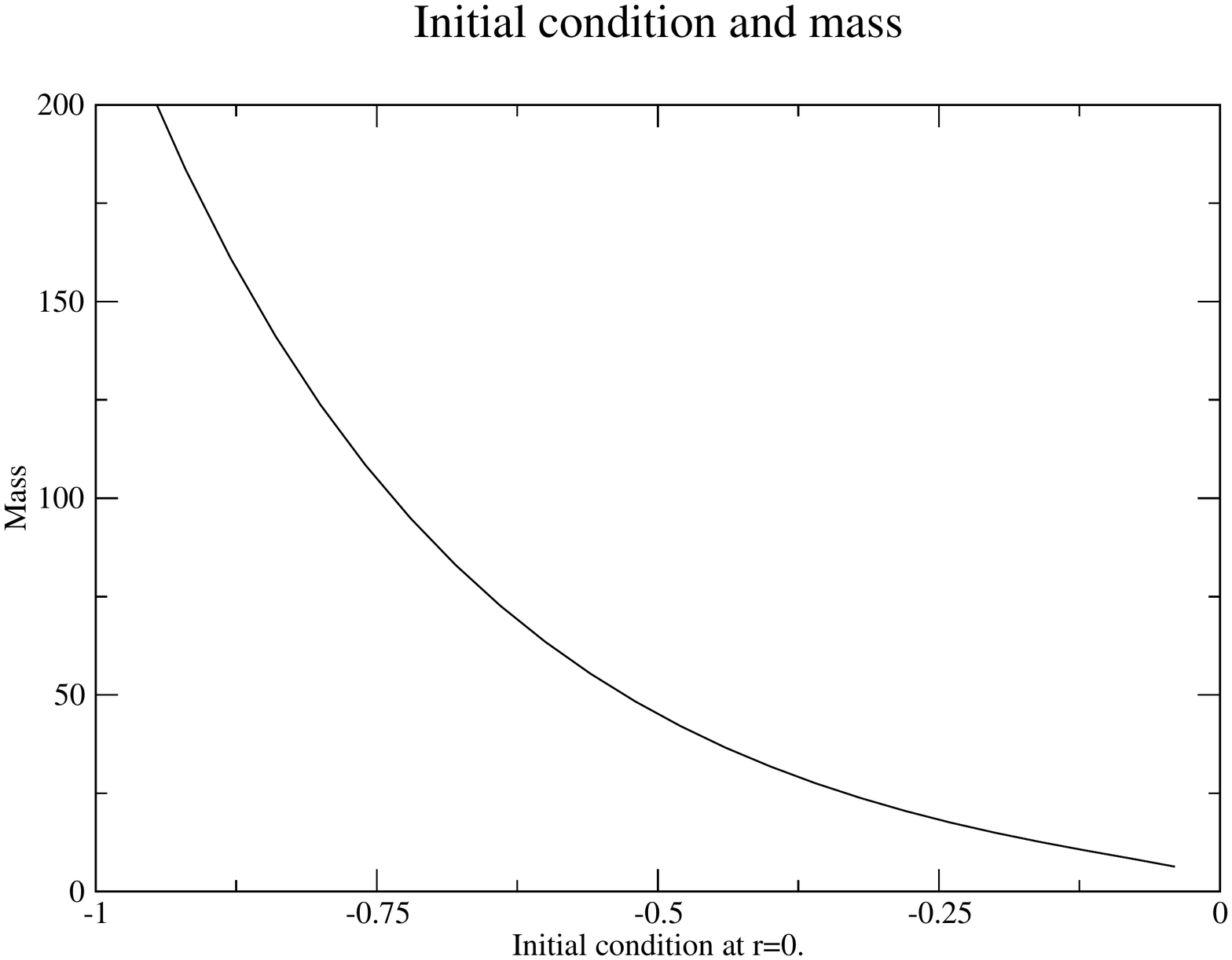}}
\caption{Masses $\widetilde{f}$ associated to the solutions of (\ref{poisonradscaled})}
\label{fig2}
\end{figure}

Figure \ref{fig2} shows the relation of the mass in terms of the  
$\widetilde{\psi}_i$. We observe that 
$$
\widetilde{\psi}_1(0) < \widetilde{\psi}_2(0)
\quad \Longrightarrow \quad
\|\widetilde{f}_1\| > \| \widetilde{f}_2 \|\, .
$$
Combining both estimates we conclude that the mass of the functions $\widetilde{f}$ is 
a decreasing function of $E_0$.
The masses of the original $f$ and the scaled $\widetilde{f}$ are related by
$$
\| f\|_1 = \frac{b}{E_0} \|\widetilde{f}\|_1.
$$
Since $b/E_0$ is also a strictly decreasing function of $E_0$, we deduce that the mass  is a decreasing function of $E_0$, and  
in consequence this proves the uniqueness of minimizers.

In the remainder of this section we wish to comment on the validity of our numerical study. Let us note that the integral term in  (\ref{poisonradscaled}) can be writen equivalently 
as
\begin{equation}\label{integral}
 \int_{\R^3} \frac{\widetilde{f}}{\widetilde{E}} \,dp 
 = 
 \frac{4 \pi} {k +1}\int_0^{\sqrt{1 - e^{\psi(r)}}}
 \left(1 - \sqrt{e^{2 \widetilde{\psi}(r)} + \xi^2}\right)^{k+1} d\xi
\end{equation}
by using radial coordinates in the variable $p$ and integrating by parts. In general these 
integrals cannot be computed  explicitly, but this is possible for
$k=1$, which avoids undesirable aproximation errors in the equation. However, our numerical simulations
show that  the solutions to equation (\ref{poisonradscaled}) with initial conditions 
(\ref{ic0scaled}) typically have  the same behavior as for $k=1$, i.e., two different qualitative shapes but only one of them is 
compatible with the  constraints of the system  (\ref{icscaled}). For $k=1$, the integral (\ref{integral}) reads
\begin{eqnarray*}
\int_{\R^3} \frac{\widetilde{f}}{\widetilde{E}} \,dp \hspace{10cm}
\\=
\frac{2 \pi}{3} \left( \sqrt{1-e^{2\psi(r)}} (1-2e^{2\psi(r)})
-3 e^{2\psi(r)}  \log{
\left(\frac{1+\sqrt{1-e^{2\psi(r)}}}{e^{\psi(r)}}\right)}
\right) \,.
\end{eqnarray*}
In order to avoid the singularity in $r = 0$ exhibited by equation
(\ref{poisonradscaled}) { we have performed numerical simulations} with initial conditions \begin{equation}
\widetilde{\psi}(\epsilon) = a  \leq 0\, , \quad
\widetilde{\psi}'(\epsilon) = 0,
\mbox{ where } \epsilon = 10^{-5}
\end{equation}
which is a reasonable aproximation of initial conditions (\ref{ic0scaled}), due to the vanishing of $\widetilde{\psi}'$ at $r= 0$.
The mass of the functions $\widetilde{f}$ can be easily computed once $\widetilde{\psi}'$ is known.

A final interesting remark about these simulations is the validation of the virial relation for any computed pair $\widetilde{f},\ \widetilde{\psi}$. Let us observe that the scaling deriving $\widetilde{f},\ \widetilde{\psi}$ from $f,\ \psi$ affects in the same way both terms in the virial relation.
Since these functions verify
\begin{equation}
\int_{\R^3}  \frac{|p|^2}{E} \widetilde{f}(x,p)\,dx = \frac{3}{k+1} \int_{\R^3} \widetilde{f}^{1+1/k}(x,p) \,dx \quad \forall p \in \R^3 \label{remark1} \end{equation} and \begin{equation} \int_{\R^3}| \nabla \widetilde{\phi} |^2 \, dx = \int_{\R^3} -\widetilde{\phi} \int_{\R^3} \frac{\widetilde{f}}{\widetilde{E}} \,\, dx \label{condition} \end{equation} where $-\widetilde{\phi}(x) =\widetilde{\psi}(|x|)$, the radial versions of these expressions allow to compute (by means of  integrals over finite intervals) both terms.
The results obtained in our simulations show a nearly total agreement (errors of order $10^{-2}$) between both terms for each solution, indicating that these functions verify in general the Virial Theorem.

\vspace{0.5cm}

{\bf Acknowledgement}:
This work
was partially supported by  MEC (Spain), Project MTM2005-02446 and by Junta de Andaluc\'{\i}a, Project E-792. Also, S.C. acknow\-ledges support
by the FCT, Portugal, contract SFRH/BDP/\-21001/2004.



\end{document}